\documentstyle[12pt]{article}
\newcommand{\nc}{\newcommand}
\nc{\bd}{
\begin{document}}
\nc{\ed}{\end{document}}
\nc{\be}{\begin{equation}}
\nc{\ee}{\end{equation}}
\nc{\ba}{\begin{eqnarray}}
\nc{\ea}{\end{eqnarray}}
\nc{\bfig}{\begin{figure}}
\nc{\efig}{\end{figure}}
\nc{\btab}{\begin{table}}
\nc{\etab}{\end{table}}

\nc{\nn}{\nonumber}
\nc{\ul}{\underline}
\nc{\ra}{\rightarrow}
\nc{\la}{\leftarrow}
\nc{\longra}{\longrightarrow}
\nc{\longla}{\longleftarrow}
\nc{\Ra}{\Rightarrow}
\nc{\La}{\Leftarrow}
\nc{\Longra}{\Longrightarrow}
\nc{\Longla}{\Longleftarrow}
\nc{\lra}{\leftrightarrow}
\nc{\Lra}{\Leftrightarrow}
\nc{\longlra}{\longleftrightarrow}
\nc{\Longlra}{\Longleftrightarrow}
\nc{\eps}{\epsilon }
\nc{\sig}{\sigma}
\nc{\sigmunu}{\sigma_{\mu\nu}}


\nc{\id}{{\bf{1}}}
\nc{\C}{{\sf C\hspace*{-0.90ex}\rule{0.15ex}{1.5ex}\hspace*{0.9ex}}}
\nc{\N}{{\sf N\hspace*{-1.05ex}\rule{0.15ex}{1.1ex}\hspace*{1.0ex}}}
\nc{\Q}{{\sf Q\hspace*{-1.15ex}\rule{0.15ex}{1.5ex}\hspace*{1.1ex}}}
\nc{\R}{{\sf R\hspace*{-1.15ex}R\hspace*{0.2ex}}}
\nc{\Z}{{\sf Z\hspace*{-1.00ex}Z\hspace*{0.2ex}}}

\def\spose#1{\hbox to 0pt{#1\hss}}
\def\ltappr{\mathrel{\spose{\lower 3pt\hbox{$\mathchar"218$}}
\raise 2.0pt\hbox{$\mathchar"13C$}}}
\def\gtappr{\mathrel{\spose{\lower 3pt\hbox{$\mathchar"218$}}
\raise 2.0pt\hbox{$\mathchar"13E$}}}
\def\qed{\hbox{\hskip 6pt\vrule width6pt height7pt depth1pt\hskip1pt}}


\def\vev#1{\left\langle #1 \right\rangle}
\def\Vev#1{\biggl\langle #1 \biggr\rangle}
\def\bra#1{\left\langle #1\right|}
\def\ket#1{\left| #1\right\rangle}
\def\Bra#1{\biggl\langle #1\biggr|}
\def\Ket#1{\biggl| #1\biggr\rangle}
\def\abs#1{\left| #1\right|}

\def\tr{{\rm tr}}
\def\Tr{{\rm Tr}}
\def\e{{\rm e}}
\def\i{{\rm i}}
\def\asinh{{\rm asinh}}
\def\acosh{{\rm acosh}}
\def\atanh{{\rm atanh}}

\nc{\half}{\frac{\,_1}{\,^2}}           
\nc{\sfrac}[2]{\frac{\,_{#1}}{\,^{#2}}} 
\nc{\nil}{\mbox{$\not\!o\ $}}           
\nc{\dslash}{\mbox{$\not{\hspace{-0.35ex}\partial}$}} 
\nc{\Dslash}{\mbox{$\not{\hspace{-0.75ex}D}$}}        
\nc{\vslash}{\mbox{$\not{\hspace{-0.40ex}v}$}}        
\nc{\pslash}{\mbox{$\not{\hspace{-0.45ex}p}$}}        

\def\gev{{\rm GeV}}
\def\tev{{\rm TeV}}
\def\mev{{\rm MeV}}
\def\kev{{\rm KeV}}
\def\ev{{\rm eV}}
\def\cm{{\rm cm}}


\nc{\CO}{{\cal O}}
\nc{\op}{{\cal O}}
\nc{\bxv}[1]{\biggl\langle #1 \biggr\rangle}  	
\nc{\g}{\gamma}
\nc{\x}{\xi}
\nc{\gnu}{\gamma_\nu}
\nc{\gmu}{\gamma_\mu}
\nc{\xnu}{\xi_\nu}
\nc{\xmu}{\xi_\mu}
\nc{\gs}{\gamma_S}
\nc{\xf}{\xi_F}
\nc{\gsp}{\gamma_{S'}}
\nc{\xfp}{\xi_{F'}}
\renewcommand{\o}[3]{\overline{(#1 \otimes #2)}_{#3}}
\nc{\oo}[3]{\overline\overline{(#1 \otimes #2)}_{#3}}

\nc{\chib}{\overline{\chi}}
\nc{\chid}{\chi_{\rm d}}
\nc{\chis}{\chi_{\rm s}}
\nc{\chidp}{\chi_{\rm d'}}
\nc{\chisp}{\chi_{\rm s'}}
\nc{\chibd}{\chib_{\rm d}}
\nc{\chibs}{\chib_{\rm s}}
\nc{\chibdp}{\chib_{\rm d'}}
\nc{\chibsp}{\chib_{\rm s'}}
\nc{\ckk}{C_{KK'}}
\nc{\ck }{C_{K }}
\nc{\ckp}{C_{K'}}
\nc{\akk}{{\cal A}_{KK'}}
\nc{\ak }{{\cal A}_{K }}
\nc{\akp}{{\cal A}_{K'}}
\nc{\fckk}{\widetilde{C}_{KK'}}
\nc{\fck }{\widetilde{C}_{K }}
\nc{\fckp}{\widetilde{C}_{K'}}


%
%
\title{One Spin Trace Formalism for $B_K$}
\author{Weonjong Lee
\thanks{Research sponsored in part by the U.S. Department of
Energy.}
and Markus Klomfass\\
{\small \it Department of Physics,
Pupin Physics Laboratories,}\\
{\small \it Columbia University,
New York, N.Y. 10027} }
\date{\today}
\bd
\begin{titlepage}
\begin{flushright}
CU-TP-642
\end{flushright}
\vspace*{\fill}
\vspace{.25in}

\begin{center}

{\Large\bf One Spin Trace Formalism for $B_K$\\}
\vspace*{.25in}
Weonjong Lee \footnote{\tt e-mail: wlee@cuphyg.phys.columbia.edu}
and Markus Klomfass\\
\vspace*{.15in}
{\small \it Department of Physics, Pupin Physics Laboratories,
Columbia University, New York, NY 10027}\\
\vspace*{.15in}
\end{center}
\vspace*{1.25in}
{\centering \today \\}
\vspace*{1.25in}
{\centering \bf ABSTRACT \\ }
It has been known for some time
that there are two methods to
calculate $ B_K $ with staggered fermions: one is the two
spin trace formalism and the other is the one spin trace
formalism. Until now, the two spin trace formalism
has been exclusively used for weak matrix element
calculations with staggered fermions.
Here, the one spin  trace formalism to calculate $ B_K $
with staggered fermions is explained. It is shown that
the one spin trace operators require
additional chiral partner
operators in order to keep the continuum chiral
behavior. The renormalization of the one
spin trace operators is described and compared
with the two spin trace formalism.
\vspace*{\fill}
\end{titlepage}
\maketitle
\begin{abstract}
It has been known for some time
that there are two methods to
calculate $ B_K $ with staggered fermions: one is the two
spin trace formalism and the other is the one spin trace
formalism. Until now, the two spin trace formalism
has been exclusively used for weak matrix element
calculations with staggered fermions.
Here, the one spin  trace formalism to calculate $ B_K $
with staggered fermions is explained. It is shown that
the one spin trace operators require
additional chiral partner
operators in order to keep the continuum chiral
behavior. The renormalization of the one
spin trace operators is described and compared
with the two spin trace formalism.
\end{abstract}
\section{Introduction}

{}From the Standard Model,
one can derive the low energy effective
Hamiltonian of electro-weak interactions
by decoupling heavy particles such as
the $ W^{\pm}$, $Z$ bosons and the  t, b, c quarks
\cite{b-lee,wise1,wise2,wise3,paschos}.
The resulting effective
Hamiltonian is composed of four-fermion operators
and bilinear operators.

In order to deduce the CP violation phase in the
CKM matrix from the measured
weak parameter $ \epsilon $,
we require the Kaon B
parameter
\cite{b-lee,wise1,paschos}.
To extract $ B_K $, we need to know
the weak matrix element
$ {\cal M}_K $ defined as
\ba
{\cal M}_K = \langle \bar{K}^{0} \mid [\bar{s} \gamma_{\mu}
(1 + \gamma_{5}) d][\bar{s} \gamma_{\mu}(1 + \gamma_{5}) d]
\mid K^{0} \rangle \ .
\ea
$ B_K $ is the ratio of $ {\cal M}_K $ over its vacuum
saturation value,
\ba
B_K =  \frac{ {\cal M}_K }
{\frac{8}{3}
\langle \bar{K}^{0} \mid [\bar{s} \gamma_{\mu}\gamma_{5} d]
\mid 0 \rangle
\langle 0 \mid [\bar{s} \gamma_{\mu}\gamma_{5} d]
\mid K^{0} \rangle } \ .
\ea
In order to calculate $ {\cal M}_K $ on the lattice,
we need to find a set of operators which can describe
the same physics on the lattice as the continuum
four-fermion operator.

Especially for the weak matrix elements
involving the pseudo-Goldstone
bosons (for example pions, kaons, etc),
it has been preferred
to take maximal advantage of
the $ U_A(1) $ symmetry of
the staggered-fermion action
which is not manifest in the lattice
Wilson fermion action\cite{sharpe0,sharpe1}.
This is the reason
why we choose staggered fermions
for our weak matrix element discussion in this article.

There are two methods to transcribe a continuum weak
matrix element (for example $ B_K $)
onto the lattice \cite{sharpe0}:
one is the one spin trace formalism and
the other is the two spin trace formalism.
The four fermion operators
can be expressed as products
of operators bilinear in the fermion fields.
In the one spin trace formalism each external hadron is
contracted with both bilinears of the four-fermion operators
simultaneously. In the two spin trace formalism
each external hadron is contracted with only
one of the bilinears in the four-fermion operators
\cite{sharpe0,sharpe1}.
By Fierz rearrangement, these two choices describe
the same continuum physics.

Until now,
the two spin trace formalism (2TR)
has been used exclusively
for calculations of weak matrix elements
with staggered fermions
in lattice QCD simulations
\cite{sharpe0,sharpe1,fuku}.
In this article, we would like to explain
the {\em one spin trace formalism} (1TR) which can
also be used to calculate
weak matrix elements (for example $ B_K $)
with staggered fermions \cite{sharpe1,kilcup0}.

Let us summarize the main contents in this article.
There is a difficulty in transcribing the one spin trace
operator to the lattice
in a way which preserves the same chiral behavior and
the same leading logarithmic behavior as the continuum
$ \Delta S = 2 $ operator.
The key point is that we must add
specific operators named
{\em chiral partner operators} to each channel
in order to keep the correct chiral behavior.
The question then is whether the additional chiral
partner operators can still guarantee the same
leading logarithmic behavior as the continuum operator.
It is shown that the answer is {\em yes}.
For a specific representation of the
exact staggered fermion chiral symmetry group
$ U_A(1) $, it is shown that both
correct chiral behavior and correct leading logarithmic
behavior are guaranteed.
The chiral behavior and the one-loop renormalization of the
operators in both formalisms will be compared with each other.
In order to make the arguments more clear and specific,
we will restrict our discussion
to $ B_K $.
Most of the arguments can be extended simply
to other weak matrix element calculations
in the one spin trace formalism
on the lattice.

This paper is organized as follows:
in section 2, we introduce the notation for
staggered fermions, definitions for bilinear
and four-fermion operators and the Fierz transformation
of operators on the lattice.
In section 3, we explain how to transcribe
operators for $ B_K $ both in the two spin trace formalism
and in the one spin trace formalism.
In section 4, we derive the chiral Ward identities
and the chiral limit of the continuum and lattice
operators. We also introduce the concept of
chiral partner operators
for the lattice operators
in the one spin trace formalism.
In section 5, we carry out the renormalization for the
bilinear and four-fermion operators. Then
we connect the lattice operators with the continuum
operators at the one loop level. We also explain the
effect of the chiral partner operators on the
renormalization of the one spin trace operators.
In the end, we give a summary and conclusions.
\section{Notations and Terminology}
In this section we will specify
our notation for the action,
fermion fields and composite operators on the lattice.
Also we will introduce the concepts of
one spin trace operators and
two spin trace operators.
We will explain the difference in the
Fierz transformations of the lattice
and continuum four fermion operators.
\subsection{Action and Operators}
The action for the staggered fermions in the notation
used in Ref. \cite{smit1} is
\ba
S = a^4 \sum_{n} \left[ \frac{1}{2a}
\sum_{\mu} \eta_{\mu}(n)
\left(
\bar{\chi}(n) U_{\mu}(n) \chi(n + \hat{\mu}) -
\bar{\chi}(n + \hat{\mu}) U^{\dagger}_{\mu}(n) \chi(n)
\right)
+ m \bar{\chi}(n)\chi(n) \right] \ ,
\ea
where $ n = (n_1,n_2,n_3,n_4)$ is
the lattice coordinate and
$ \eta_{\mu}(n) = (-1)^{n_1 + \cdots + n_{\mu-1}} $.
The single-component staggered fermion field $ \chi $
corresponds to 4 degenerate
fermions ({\em cf.} fermion doubling problem).
Golterman and Smit proposed one method to
express the staggered fermion field
as 4 degenerate flavors
in terms of Euclidean Dirac spin and flavor matrices
obtained by chopping momentum space into 16 divisions
\cite{smit1}.
Kluberg-Stern, Morel, Napoly and Petersson
suggested that we can take hypercubes in the
coordinate space as the basic units
which correspond to the individual points
of a lattice two times coarser
and interpret the 16 hypercube coordinates
as 4 Dirac spin and 4 flavor components
\cite{klubergstern1}.
In Ref. \cite{klubergstern1},
Kluberg-Stern {\em et al.} defined the Dirac field $ Q(y) $ as
\ba
Q^{\alpha i}(y) =
\frac{1}{ N_f \sqrt{N_f}}
\sum_{A} (\gamma_{A})^{\alpha i}
\chi(y_A)
\ea
where $ \alpha $ is the Dirac spin index,
$ i $ is the flavor index and $ N_f $ is the
number of degenerate flavors ($ N_f = 4 $).
Define also
\ba
y_A \equiv 2 y + A,\ \ \mbox{with }  A \in \{0,1\}^{4}
\ea
and
\ba
\gamma_A = \gamma^{A_1}_1
\gamma^{A_2}_2 \gamma^{A_3}_3
\gamma^{A_4}_4 \ .
\ea
If we set the gauge links equal to unity,
the staggered fermion
action can be expressed in terms of the $ Q $ field
as follows:
\ba
S & = &
a^4 \sum_{y,\mu} \sum_{A,B}
[ \bar{\chi}(y_A)
( \overline{\gamma_{\mu} \otimes I } )_{AB}
\left(
\frac{\chi( (y + \hat{\mu})_B ) - \chi( (y - \hat{\mu})_B )}{4a}
\right) ]
\nonumber
\\
& &
+ a^4 \sum_{y} m \sum_{A,B}
[ \bar{\chi}(y_A)
(\overline{I \otimes I } )_{AB} \chi(y_B) ]
\nonumber
\\
& &
+ a^4 \sum_{y,\mu} \sum_{A,B}
[ \bar{\chi}(y_A)
( \overline{ \gamma_{5} \otimes \xi_{\mu5} } )_{AB}
\left(
\frac{\chi( (y + \hat{\mu})_{B})
+ \chi( (y - \hat{\mu})_B ) -2 \chi(y_B) }
{(2a)^2}
\right)]
\\
& = & (2a)^4 \sum_{y,\mu} [
\bar{Q}(y) (\gamma_{\mu} \otimes I)
\left(\frac{Q(y+\hat{\mu}) - Q(y-\hat{\mu})}{4a}
\right)] +
(2a)^4 m \sum_{y}[ \bar{Q}(y) (I \otimes I) Q(y)]
\nonumber
\\
& &
+(2a)^4 \sum_{y,\mu} [ \bar{Q}(y)
(\gamma_{5} \otimes \xi_{\mu5} )
\left(
\frac{Q(y + \hat{\mu}) + Q(y - \hat{\mu})- 2 Q(y)}
{(2a)^2}
\right)]
\ea
where
$ \hat{\mu} $ represents a unit displacement on the
coarse lattice and the matrices $ \gamma_S \otimes \xi_F $
represent the standard tensors (or direct products).
The matrices $ \overline{\gamma_S \otimes \xi_F } $
are given by
\ba
( \overline{ \gamma_S \otimes \xi_F})_{AB}
\equiv \frac{1}{4}
\mbox{Tr}(\gamma^{\dagger}_A
\gamma_S \gamma_B
\gamma^{\dagger}_F)\ .
\ea
These two matrices are related to each other by
\ba
(\gamma_S \otimes \xi_F)^{\alpha i , \beta j} =
\frac{1}{4} \sum_{A,B}
\gamma_{A}^{\alpha i}
(\overline{ \gamma_S \otimes \xi_F })
\gamma_B^{\beta j} \  ,
\ea
where $ \alpha $, $ \beta $ are Dirac spin indices
and $i$, $j$ are flavor indices.
This interpretation is called {\em coordinate-space method}.

It is known that the flavor interpretation
in the coordinate-space method by Kluberg-Stern
{\em et al.} can be related to
the momentum-space method by Golterman
and Smit \cite{kieu1} as follows:
\ba
\label{k-gs}
(\overline{\overline{ \gamma_S \otimes \xi_F }})_{AB}
=
\sum_{CD} \frac{1}{N_f}(-1)^{A \cdot C}
(\overline{ \gamma_S \otimes \xi_F })_{CD}
\frac{1}{N_f}(-1)^{D \cdot B} \ .
\ea
The momentum-space matrices
$ (\overline{\overline{ \gamma_S \otimes \xi_F }})_{AB} $
introduced by Golterman and Smit
are related unitarily
to the coordinate-space matrices
$ (\gamma_S \otimes \xi_F)^{\alpha i , \beta j} $\cite{kieu1}.
Here we adopt the coordinate-space method.

The continuum limit of the staggered fermion action
on the lattice is a kind of QCD
with four degenerate flavors ($ N_f = 4 $) \cite{smit1}.
In the continuum, operators of the form
\ba
\bar{q}(x) (\gamma_S \otimes \xi_F) q(x)
\ea
can be used as interpolating fields
for fermion bilinear objects \cite{sharpe0,sharpe1},
where $ \gamma_S $ is the Dirac spin matrix and
$ \xi_{F} $ belongs to SU(4) flavor symmetry
group. There are a lot of choices to transcribe the
lattice operator for a given continuum operator.
The conventional choice of
bilinear operator which we will follow here,
is
\ba
\label{bilinear}
{\cal O}_{SF}(y) = \frac{1}{N_f}
\sum_{AB}\bar{\chi}(y_A)
( \overline{\gamma_S \otimes \xi_F } )_{AB}
{\cal U}(y_A,y_B) \chi(y_B),
\ea
where $ {\cal U}(y_A,y_B) $
is a product of gauge links
that makes $ {\cal O}_{SF} $ gauge-invariant
\cite{sheard0,sharpe1,sheard1}.
There are two simple choices for
$ {\cal U}(y_A,y_B) $.
One is that $ {\cal U}(y_A,y_B) $ is replaced
by the identity and the spinors $ \bar{\chi}(y_A) $ and
$ \chi(y_B) $ are evaluated in Landau gauge
\cite{sharpe1}
(so called {\em Landau gauge operator}).
The other is that the link matrices are
included and that $ {\cal O}_{SF} $ is made as symmetric
as possible by averaging
over all of the shortest paths between
$ y_A $ and $ y_B $ \cite{sheard0,sheard1}:
\ba
\label{ulink}
{\cal U}(y_A,y_B) & = & \frac{1}{4!}
\sum_{P} U(y_A, y_A+\Delta_{P1})\cdots
\nonumber
\\
& &
U(y_A+\Delta_{P1}+\Delta_{P2}+\Delta_{P3}\ ,\ y_B) \ ,
\ea
where
$ P $ is an element of the permutation group (1234) and
\ba
\label{delta}
\Delta_{\mu} = (B_{\mu}-A_{\mu}) \hat{\mu} \ .
\ea
There are two kinds of four-fermion operators
which have different color contraction structure.
The general form of color two-trace operators is
\be
[\bar{\chi}^{a}_{f_1} (y_A)
( \overline{\gamma_S \otimes \xi_F } )_{AB}
\chi^{b}_{f_2} (y_B)]
[\bar{\chi}^{c}_{f_3} (y_C)
( \overline{\gamma_{S'} \otimes \xi_{F'} } )_{CD}
\chi^{d}_{f_4} (y_D)]
{\cal U}_{ab}(y_A,y_B)
{\cal U}_{cd}(y_C,y_D)
\ee
and the general form of color one-trace operators is
\be
[\bar{\chi}^{a}_{f_1} (y_A)
( \overline{\gamma_S \otimes \xi_F } )_{AB}
\chi^{b}_{f_2} (y_B)]
[\bar{\chi}^{c}_{f_3} (y_C)
( \overline{\gamma_{S'} \otimes \xi_{F'} } )_{CD}
\chi^{d}_{f_4} (y_D)]
{\cal U}_{ad}(y_A,y_D)
{\cal U}_{cb}(y_C,y_B)
\ee
where $f_1$, $f_2$, $f_3$ and $f_4$
label the continuum
flavor (for example, u, d ,s, $ \cdots $) and
the two choices for $ {\cal U}_{ab}(y_A,y_B) $ are
the same as given
for the bilinear operators.
The reader should recognize that at this point we
have introduced (following Sharpe \& Kilcup
\cite{sharpe0,sharpe1}) flavor in two ways: First
the 4 degenerate flavors associated with each
staggered fermion field $ \chi(y_A) $,
labeled by the index $A$,
and second the possibly non-degenerate additional flavors
corresponding to distinct staggered fields $ \chi_f $,
labeled by the indices $ f $.

We can represent the four fermion operators
in $ B_K $ in terms of
$ S $ (scalar), $ V $ (vector) , $ T $ (tensor),
$ A $ (axial) and $ P $ (pseudo-scalar) as follows:
\ba
\label{note_1}
[V \times S]^{f_1f_2;f_3f_4}_{c_1c_2;c_3c_4}
& \equiv &
[\bar{\chi}_{f_1}^{c_1}
(\overline{ \gamma_{\mu} \otimes I } )
\chi_{f_2}^{c_2}]
[\bar{\chi}_{f_3}^{c_3}
(\overline{ \gamma_{\mu} \otimes I } )
\chi_{f_4}^{c_4}] \ ,
\\
\label{note_2}
[V \times P]^{f_1f_2;f_3f_4}_{c_1c_2;c_3c_4}
& \equiv &
[\bar{\chi}_{f_1}^{c_1}
(\overline{ \gamma_{\mu} \otimes \xi_5 } )
\chi_{f_2}^{c_2}]
[\bar{\chi}_{f_3}^{c_3}
(\overline{ \gamma_{\mu} \otimes \xi_5 } )
\chi_{f_4}^{c_4}]
\\
\label{note_3}
[A \times S]^{f_1 f_2;f_3 f_4}_{c_1 c_2;c_3 c_4}
& \equiv &
[\bar{\chi}_{f_1}^{c_1}
(\overline{ \gamma_{\mu5} \otimes I} )
\chi_{f_2}^{c_2}]
[\bar{\chi}_{f_3}^{c_3}
(\overline{ \gamma_{\mu5} \otimes I} )
\chi_{f_4}^{c_4}]
\\
\label{note_4}
[A \times P]^{f_1 f_2;f_3 f_4}_{c_1 c_2;c_3 c_4}
& \equiv  &
[\bar{\chi}_{f_1}^{c_1}
(\overline{ \gamma_{\mu5} \otimes \xi_5 } )
\chi_{f_2}^{c_2} ]
[\bar{\chi}_{f_3}^{c_3}
(\overline{ \gamma_{\mu5} \otimes \xi_5 } )
\chi_{f_4}^{c_4} ]
\ea
where $ f_i $ ($ 1 \leq i \leq 4 $)
represent the continuum flavor (u, d, s, $\cdots$),
$ c_i $ ($ 1 \leq i \leq 4 $) represent the color
indices in the fundamental representation of SU(3)
and  for  example
$ [A \times P]^{f_1f_2;f_3f_4}_{c_1c_2;c_3c_4} $
represents a four-fermion operator
on the lattice
which has the axial spin structure $A$ and the pseudoscalar-like
flavor structure $P$.
\subsection{Fierz Transformation}
Let us introduce the following
notation \cite{sharpe1} :
\ba
& & S = I \otimes I,\
V = \sum_{\mu} \gamma_{\mu} \otimes \gamma_{\mu}, \
T = \sum_{\mu<\nu} \sigma_{\mu\nu} \otimes \sigma_{\mu\nu},
\nonumber
\\
& &
\label{notation}
A = \sum_{\mu} \gamma_{\mu5} \otimes \gamma_{\mu5},\
P = \gamma_{5} \otimes \gamma_{5}
\ea
where $ \otimes $ means direct product and
$ \sigma_{\mu\nu} = \frac{1}{2}
[\gamma_{\mu},\gamma_{\nu}]$.
The same notation also applies to the flavor structure
$ \xi_F \otimes \xi_F $ by switching $ \gamma $ to $ \xi $.

In terms of the notation given in Eq. (\ref{notation}),
the Fierz transformation of $ S $, $ V $, $ T $, $ A $
and $ P $
can be expressed as follows:
\ba
& &
(\Gamma_1, \Gamma_2, \Gamma_3, \Gamma_4, \Gamma_5)
\equiv  (S, V, T, A, P)
\nonumber \\
& &
(\Gamma_i )_{\alpha,\beta';\alpha',\beta}
= F_{ij} \ (\Gamma_j )_{\alpha,\beta;\alpha',\beta'}
\label{fierz1}
\\
& &
\left[ F_{ij} \right]
 =
\frac{1}{4}
\left[
\begin{array}{ccccc}
  1	&  1	&  -1	&  -1	&  1 \\
  4	&  -2	&  0	&  -2	&  -4 \\
  -6	&  0	&  -2	&  0	&  -6  \\
  -4	&  -2	&  0	&  -2	&  4  \\
  1	&  -1	&  -1	&  1	&  1
\end{array}
\right]
\ea
{}From this relationship, we can show that the following
two operators are identical:
\ba
\bar{S}^{a} \gamma_{\mu}(1-\gamma_5) D^{a}
\bar{S}^{b} \gamma_{\mu}(1-\gamma_5) D^{b}
=
\bar{S}^{a} \gamma_{\mu}(1-\gamma_5) D^{b}
\bar{S}^{b} \gamma_{\mu}(1-\gamma_5) D^{a}.
\ea
The continuum $ \Delta S = 2 $ operator is
Fierz-transformed into itself.

Let us use the above relationship
(Eq. (\ref{fierz1})) to obtain the Fierz
transformation of the lattice operators for staggered fermions.
In terms of $ Q^{\alpha,i} $, the spin matrix and flavor
matrix can be separated completely.
Therefore, the spin and flavor matrices are Fierz-transformed
separately according to the relationship (Eq. (\ref{fierz1})).
The Fierz transformations
for the lattice four-fermion operators are
\ba
& &
\mbox{for } \
\Gamma_i,\Gamma_j, \Gamma_l, \Gamma_k \in \{ S,V,T,A,P \}
\nonumber \\
& &
\label{fierz2}
[\Gamma_i \times \Gamma_j]^{f_1,f_4;f_3,f_2}_{c_1,c_4;c_3,c_2}
= - F_{il} \ F_{jk} \
[\Gamma_l \times \Gamma_k]^{f_1,f_2;f_3,f_4}_{c_1,c_2;c_3,c_4}
\ea
where $ \Gamma_i $, $ \Gamma_l $ represent the spin structure
and $ \Gamma_j $, $ \Gamma_k $ represent the flavor structure
as explained in Eq. (\ref{note_1}-\ref{note_4}).

Now let us obtain some useful relations which
will
important later. From the lattice Fierz transformations
Eq.(\ref{fierz2}), we obtain the following relationship
on the lattice:
\ba
& \frac{1}{2} &
[V \times S + A \times P
+ V \times P + A \times S
]^{f_1,f_4;f_3,f_2}_{c_1,c_4;c_3,c_2}
\nonumber
\\
& &= \frac{1}{N_f}[ (V + A) \times (S - T + P)
]^{f_1,f_2;f_3,f_4}_{c_1,c_2;c_3,c_4}
\ea
which  will be used later to show that
the lattice operator in the  one spin trace formalism is
different from the corresponding operator
in the two spin trace formalism, as well as to show that
the Fierz transform property of the continuum $ B_K $
will be recovered in the continuum limit of
$ a = 0 $. This is because
the operators with different
flavor structure from that ($\xi_5$) of the pseudo-Goldstone pion,
($ [ (V+A) \times S ] $ and $ [ (V+A) \times T ] $)
can not contribute to $ B_K $
in the continuum limit of $ a = 0 $.
\section{Operator Transcription for $ B_K $}
In this section we explain how to choose the lattice
operator in order to calculate the weak matrix element
found in $ B_K $.
We follow quite closely the original
work on the two spin trace formalism of
transcribing operators for the weak matrix elements
done in Ref. \cite{sharpe1,daniel1}.
In this section,  we explain and compare
two spin trace formalism and one spin trace formalism
in detail.

The continuum operator related to $ B_K $
\cite{wise1,paschos,b-lee} is
\ba
\label{deltas2}
O^{Cont}_{\Delta S = 2} & = &
[\bar{s}^{a} \gamma_{\mu}(1 - \gamma_{5}) d^{a}]
[\bar{s}^{b} \gamma_{\mu}(1 - \gamma_{5}) d^{b}]
\\
& = &
[\bar{s}^{a} \gamma_{\mu}(1 - \gamma_{5}) d^{b}]
[\bar{s}^{b} \gamma_{\mu}(1 - \gamma_{5}) d^{a}]\ .
\ea
As you can see from the above equations,
the matrix element in $ B_K $ is
Fierz-transformed into itself in the continuum.
Since there are four degenerate flavors on the lattice,
the Fierz transformation is completely different on the
lattice from that in the continuum.
This difference allows two different ways of
transcribing operators on the lattice.

The continuum $ \Delta S = 2 $ operator
has only two flavors ($ s $ and $ b $).
Following Ref. \cite{sharpe0,sharpe1},
we may also introduce four valence quark flavors
($ S $, $ S'$ and $ D $, $D'$)
to represent the same physics:
\ba
{\cal M}_K & = &
\langle \bar{K}^{0} \mid
[\bar{s}^{a} \gamma_{\mu}(1 + \gamma_{5}) d^{a}]
[\bar{s}^{b} \gamma_{\mu}(1 + \gamma_{5}) d^{b}]
\mid K^{0} \rangle
\\
\longrightarrow
{\cal M}_K & = &
2 \langle \bar{K}^{0} \mid
[\bar{S}^{a} \gamma_{\mu}(1 + \gamma_{5}) D^{a}]
[\bar{S'}^{b} \gamma_{\mu}(1 + \gamma_{5}) D'^{b}]
\mid K'^{0} \rangle
\nonumber
\\
\label{cont2}
& &+
2 \langle \bar{K}^{0} \mid
[\bar{S}^{a} \gamma_{\mu}(1 + \gamma_{5}) D^{b}]
[\bar{S'}^{b} \gamma_{\mu}(1 + \gamma_{5}) D'^{a}]
\mid K'^{0} \rangle
\\
\mbox{or } \longrightarrow
{\cal M}_K & = &
2 \langle \bar{K}^{0} \mid
[\bar{S}^{a} \gamma_{\mu}(1 + \gamma_{5}) D'^{a}]
[\bar{S'}^{b} \gamma_{\mu}(1 + \gamma_{5}) D^{b}]
\mid K'^{0} \rangle
\nonumber
\\
\label{cont1}
& & +
2 \langle \bar{K}^{0} \mid
[\bar{S}^{a} \gamma_{\mu}(1 + \gamma_{5}) D'^{b}]
[\bar{S'}^{b} \gamma_{\mu}(1 + \gamma_{5}) D^{a}]
\mid K'^{0} \rangle
\ea
where the primed and unprimed flavors
are supposed to have the same mass.
We have introduced additional species $ S'$ and $ D'$
to indicate a particular contraction.
Similarly, the hadronic eigenstate $ K^0 $ is composed of the
valence quarks $ S $ and $ D $,
while $ K'^0 $ contains $ S' $ and $ D' $.
In the Eq. (\ref{cont2}),
the first bilinear has to be contracted with $K$
and the second bilinear with $K'$.
This is the so-called {\it two spin trace} contraction.
Another possibility is the {\it one spin trace} contraction
as in the Eq. (\ref{cont1}) where
the external kaon is contracted with both bilinears.

As you can see in the above, the Fierz transformation
connects Eq.(\ref{cont2}) with Eq.(\ref{cont1}).
They are identical to each other.
But the lattice version
of Eq.(\ref{cont2}) and Eq.(\ref{cont1})
can not be related by Fierz transformation.
Thus, the righthand side of Eq.(\ref{cont2}) corresponds
to two spin trace formulation
of the operator on the lattice,
while the righthand side of Eq.(\ref{cont1}) corresponds
to one spin trace formulation,
as will be described later.
\subsection{Two Spin Trace Formalism for $ B_K $}

The detailed explanation
of two spin trace formalism for $ B_K $
is given in Ref. \cite{sharpe0,sharpe1}.

First of all we need a Kaon operator on the lattice.
Since the conserved axial current has flavor $ \xi_{5} $,
it is better to use the composite operator of
the lattice pseudo-Goldstone boson
\cite{sharpe1} with
the flavor matrix $ \xi_{5} $ as a Kaon operator:
\ba
& & K(n) = \bar{\chi}_{\rm d}(n) \epsilon(n) \chi_{\rm s}(n)
\nn \\
& & K(t) \equiv \sum_{\vec{n}} K(\vec{n},t)
\\
& & K'(n) = \bar{\chi}_{\rm d'}(n) \epsilon(n) \chi_{\rm s'}(n)
\nn \\
& & K'(t) \equiv \sum_{\vec{n}} K'(\vec{n},t)
\ea
%
%
%
%
where $ \sum_{\vec{n}} $ represents the summation over
spatial volume needed to construct an operator for
zero-momentum Kaon.
We need to specify the four fermion operator on the lattice
which corresponds to the continuum four fermion operator
in Eq.(\ref{cont2}).
The two spin trace formalism means
that the operators consist
of two bilinears and that each external kaon is
contracted with one of these bilinears.
Hence the four fermion operator is
\ba
& & O^{Latt}_{2TR} =
(V \times P)^{2TR}_{ab;ba}
+ (V \times P)^{2TR}_{aa;bb}
+ (A \times P)^{2TR}_{ab;ba}
+ (A \times P)^{2TR}_{aa;bb}
\label{op-2tr}
\ea
with
\ba
& &
(V \times P)^{2TR}_{ab;ba} \equiv
\frac{1}{N_f^2}
[V \times P]^{\rm sd;s'd'}_{ab;ba}
\label{op-2tr-1}
\\
& &
(V \times P)^{2TR}_{aa;bb} \equiv
\frac{1}{N_f^2}
[V \times P]^{\rm sd;s'd'}_{aa;bb}
\label{op-2tr-2}
\\
& &
(A \times P)^{2TR}_{ab;ba} \equiv
\frac{1}{N_f^2}
[A \times P]^{\rm sd;s'd'}_{ab;ba}
\label{op-2tr-3}
\\
& &
(A \times P)^{2TR}_{aa;bb} \equiv
\frac{1}{N_f^2}
[A \times P]^{\rm sd;s'd'}_{aa;bb}
\label{op-2tr-4}
\ ,
\ea
where $a$ and $b$ represent color indices
in the fundamental representation of $ SU(3) $ and the notations
in the righthand side are as defined
in Eqs. (\ref{note_1}-\ref{note_4}).
As you can see in the above, we choose the vector and
axial channel
in such a way  that the flavor structure is
the same as that of the pseudo-Goldstone
boson (Kaon) and the contraction with the external Kaons
has a non-vanishing flavor trace.
The operators with different flavor structure from the
pseudo-Goldstone boson (Kaon) are supposed to vanish
when it is contracted with external Goldstone Kaons,
because of vanishing flavor trace.
However, one needs to note that
the argument of vanishing flavor trace holds only for
tree-level discussion (i.e. in the classical limit),
because the radiative corrections
in lattice QCD mix an operator with operators
of  different flavor matrices. This point will become
more clear later when we discuss about the renormalization.

The two spin trace form of $ {\cal M}_K $ is prescribed by
\ba
\label{m_2tr}
{\cal M}_K^{2TR}(t,t^{'}) =
< K(t) O^{Latt}_{2TR} K'(t^{'}) > \ \ .
\ea
\subsection{One Spin Trace Formalism for $ B_K $}

For the one spin trace formalism,
we use the same operator for the kaon as
in the two spin trace formalism.
The 4-fermion operator in one spin trace form, however,
is
\ba
& & O^{Latt}_{1TR} =
(V \times P)^{1TR}_{ab;ba}
+ (V \times P)^{1TR}_{aa;bb}
+ (A \times P)^{1TR}_{ab;ba}
+ (A \times P)^{1TR}_{aa;bb}
\nn \\
& &
\label{1tr-op}
+ O^{1TR}_{ \mbox{\small chiral partner} }
\ea
with
\ba
& &
\label{1tr-1}
(V \times P)^{1TR}_{ab;ba} \equiv
\frac{1}{2N_f}
[V \times P]^{\rm sd';s'd}_{ab;ba}
\\
& &
(V \times P)^{1TR}_{aa;bb} \equiv
\frac{1}{2N_f}
[V \times P]^{\rm sd';s'd}_{aa;bb}
\\
& &
(A \times P)^{1TR}_{ab;ba} \equiv
\frac{1}{2N_f}
[A \times P]^{\rm sd';s'd}_{ab;ba}
\\
& &
\label{1tr-4}
(A \times P)^{1TR}_{aa;bb} \equiv
\frac{1}{2N_f}
[A \times P]^{\rm sd';s'd}_{aa;bb}
\ ,
\ea
where the notations in the righthand side
of the equations
are defined in Eqs. (\ref{note_1}-\ref{note_4}) and
$ O^{1TR}_{\mbox{\small chiral partner}} $
will be determined explicitly later
when we discuss the chiral behavior.
In the one spin trace case, the external kaons have to be
contracted with both bilinears simultaneously.
The one spin trace form of $ {\cal M}_K $ is defined as
\ba
\label{m_1tr}
{\cal M}_K^{1TR}(t,t^{'}) =
< K(t) O^{Latt}_{1TR} K'(t^{'}) > \ \ .
\ea
%
%

Let us explain the choices of axial and vector spin terms
in Eq. (\ref{1tr-op}).
In contrast with the two spin trace case,
for any choice of the flavor matrix $ F $ (i.e. $ \xi_F $),
the contraction of the external Goldstone
Kaon fields with the $ (V \times F) $
and $ (A \times F) $ does not vanish.
However, in order to take maximal advantage of $ U_A(1) $
symmetry, we will choose the axial channel
as $ (A \times P) $.
In order to achieve the same leading logarithmic
behavior in the one spin trace form on the lattice
as the continuum operator, the $ (A \times P) $ operator
needs $ (V \times P) $ as a vector channel.
This point will become clear later
when we discuss renormalization.

In order to achieve the same chiral behavior for the
one spin trace form on the lattice as the
continuum $ \Delta S = 2 $ operator,
it is required that
once we choose the flavor of the axial current,
we should add a vector current which
corresponds to the chiral partner
of the given axial channel
(for example, $ (A \times P) $ operator needs
$ (V \times S) $ operator as a chiral partner).
Once we add the chiral partner
to the given axial channel,
the operator in the one spin trace formalism
can have the same chiral behavior
on the lattice as in the continuum.
The point is that
for the vector and axial channel to satisfy the
appropriate chiral limit,
the correct chiral partners must be added.
\section{Chiral Behavior}
With the two lattice prescriptions for $ B_K $ introduced
in the previous section, we have to prove that each
formalism separately possesses the correct chiral behavior.
In the continuum, one can
show that $ \int dt\ dt^{'} \ {\cal M}_K(t,t^{'}) $
vanishes both in the limit of $ m_{d} = m_{s} $ and the
limit of $ m_{d^{'}} = m_{s^{'}} $.
For our proof, we will have to use several lattice
Ward identities which will be explained in the
following sections.
\subsection{Continuum Chiral Ward Identity}
In the continuum we have the following Ward identity
\ba
\lefteqn{
( m_{s} + m_{d} ) \int d^{4}y
\langle 0 \mid T [ S^{\alpha}(x)
\bar{S}^{\rho}(y) (\gamma_{5})_{\rho,\sigma}
D^{\sigma}(y)
\bar{D}^{\beta}(z) ]\mid 0 \rangle}
\nonumber
\\
\label{c_chiral1}
&=& (\gamma_{5})_{\alpha \rho}
    \langle 0 \mid T [D^{\rho}(x)
    \bar{D}^{\beta}(z)] \mid 0 \rangle
    + \langle 0 \mid T [ S^{\alpha} (x)
    \bar{S}^{\rho}(z)] \mid 0 \rangle
    (\gamma_{5})_{\rho \beta} \;,
\ea
where $ T $ indicates the time ordered product.

Using Eq.(\ref{c_chiral1}), we can
obtain the following relationship
\ba
& &(m_{d}+m_{s})(m_{d^{'}}+m_{s^{'}})
\int dt dt^{'} {\cal M}_K(t,t^{'})
\nn \\
\label{c_chiral2}
& &= \sum_{C = V,A}
\langle 0\mid
[\bar{S'} \gamma_{C} \gamma_{5} S'
-\bar{D'} \gamma_{C} \gamma_{5} D']
[\bar{S} \gamma_{C} \gamma_{5} S
-\bar{D} \gamma_{C} \gamma_{5} D]
\mid 0 \rangle \ ,
\ea
where
\ba
\label{m_cont}
{\cal M}_K(t,t') \equiv  \langle
K^{Cont}(t)
{\cal O}^{Cont}_{\Delta S = 2}
K'^{Cont}(t') \rangle
\ea
with
\ba
& &
K^{Cont}(t) \equiv
\int d^{3}\vec{y} \bar{D}(\vec{y},t)
\gamma_5  S(\vec{y},t)
\\
& &
K'^{Cont}(t) =
\int d^{3}\vec{y} \bar{D'}(\vec{y},t)
\gamma_5  S'(\vec{y},t)\ .
\ea
The above result corresponds to the chiral limit of the two
spin trace formalism.
Applying a Fierz transformation on the right-hand
side of Eq. (\ref{c_chiral2}), we obtain another form of
the Ward identity
\ba
& &(m_{d}+m_{s})(m_{d'}+m_{s'})
\int dt dt^{'} {\cal M}_K (t,t^{'})
\nn \\
& &=
\label{c_chiral3}
\sum_{C = V,A}
\langle 0 \mid
[
\bar{S'} \gamma_{C} \gamma_{5}
(S \bar{S} - D \bar{D})
\gamma_{C} \gamma_{5} S'
+
\bar{D'} \gamma_{C} \gamma_{5}
(D \bar{D} - S \bar{S})
\gamma_{C} \gamma_{5} D'
]
\mid 0 \rangle \; ,
\ea
which corresponds to the chiral limit
of the one spin trace
formalism on the lattice.
Eq. (\ref{c_chiral2}) is identical
to Eq. (\ref{c_chiral3}) in the continuum.
Since $ \bar{S} \gamma_{C} \gamma_{5} S $ and
$ \bar{D} \gamma_{C} \gamma_{5} D $ operators
cancel in the limit $ m_{\rm d} = m_{\rm s} $,
we conclude that the lefthand side must vanish as well
when $ m_{\rm d} = m_{\rm s} $.
The same is true for the limit
$ m_{\rm d'} = m_{\rm s'} $.
In the chiral limit of $ E_{\mbox{\tiny Kaon}} \rightarrow 0 $,
this behavior of
vanishing weak matrix element could also be confirmed
numerically on the lattice.
\subsection{Chiral Ward Identity of the Two Spin Trace Formalism}
Let us define the quark propagator on the lattice
as
\ba
G_{f}(x,y) \equiv \langle 0\mid
\chi_f(x) \bar{\chi}_f(y) \mid 0 \rangle \;,
\ea
%
%
where $ f $ indicates the continuum flavor.
In appendix A and Ref. \cite{tool-kit},
the following lattice version of
the continuum Ward identity Eq. (\ref{c_chiral1})
is derived
\ba
\label{ward1}
(m_d+m_s)\sum_{n}
G_{\rm d}(x,n) \epsilon(n) G_{\rm s}(n,y) =
\epsilon(x) G_{\rm d}(x,y) + G_{\rm s}(x,y) \epsilon(y) \;,
\ea
where $ n \equiv (n_x,n_y,n_z,n_t) $ and
$ \epsilon(n) \equiv (-1)^{n_x+n_y+n_z+n_t} $.
Using Eq.(\ref{ward1}),
one obtains the following
\ba
& &(m_d+m_s)(m_{d'}+m_{s'})\sum_{t,t^{'}}
{\cal M}_K^{2TR}(t,t^{'})
= \sum'_{\mu}
\nonumber \\
& &
\left\{
\langle 0 \mid
[\bar{\chi}_{\rm s'}
\overline{(\gamma_{\mu} \otimes 1 )} \chi_{\rm s'}
-\bar{\chi}_{\rm d'}
\overline{(\gamma_{\mu} \otimes 1 )} \chi_{\rm d'}]
%
%
[\bar{\chi}_{\rm s}
\overline{(\gamma_{\mu} \otimes 1 )} \chi_{\rm s}
-\bar{\chi}_{\rm d}
\overline{(\gamma_{\mu} \otimes 1 )} \chi_{\rm d}]
\mid 0 \rangle \right.
\nn \\
& &
+ \left.
\langle 0\mid
[\bar{\chi}_{\rm s'}
\overline{(\gamma_{\mu5} \otimes 1 )} \chi_{\rm s'}
-\bar{\chi}_{\rm d'}
\overline{(\gamma_{\mu5} \otimes 1 )} \chi_{\rm d'}]
\label{l_chiral1}
[\bar{\chi}_{\rm s}
\overline{(\gamma_{\mu5} \otimes 1 )} \chi_{\rm s}
-\bar{\chi}_{\rm d}
\overline{(\gamma_{\mu5} \otimes 1 )} \chi_{\rm d}]
\mid 0 \rangle
\right\}
\ea
where the details of the derivation are given
in appendix B and Ref. \cite{sharpe1}.
Equation (\ref{l_chiral1}) is quite similar
to Eq.(\ref{c_chiral2}). The sum
$ \sum' $ in Eq. (\ref{l_chiral1}) indicates that
in addition to the sum over the vector index $ \mu $
we also include both one trace and two trace color
contractions.
{}From Eq.(\ref{l_chiral1}) one can see that
the right-hand side vanishes in the limit
of $ m_d = m_s $ and also in the limit
of $ m_{d^{'}} = m_{s^{'}} $,
in accordance with
the continuum chiral behavior.
\subsection{Chiral Ward Identity
of the One Spin Trace Formalism}
In the one spin trace formalism, each channel requires
an additional set of operators with different spin-flavor
structure in order to realize the correct
chiral behavior.
{}From appendix C,
we need to choose the chiral partner operator
$ {\cal O}^{1TR}_{\mbox{\small chiral partner}} $
introduced in Eq. (\ref{1tr-op})
as follows
\ba
\label{chi-op}
{\cal O}^{1TR}_{\mbox{\small chiral partner}}
= (A \times S)^{1TR}_{ab;ba}
+ (A \times S)^{1TR}_{aa;bb}
+ (V \times S)^{1TR}_{ab;ba}
+ (V \times S)^{1TR}_{aa;bb}
\ea
with
\ba
& &
\label{chi-1}
(A \times S)^{1TR}_{ab;ba} \equiv
\frac{1}{2N_f}
[ A \times S ]^{\rm sd';s'd}_{ab;ba}
\\
& &
(A \times S)^{1TR}_{aa;bb} \equiv
\frac{1}{2N_f}
[ A \times S ]^{\rm sd';s'd}_{aa;bb}
\\
& &
(V \times S)^{1TR}_{ab;ba} \equiv
\frac{1}{2N_f}
[ V \times S ]^{\rm sd';s'd}_{ab;ba}
\\
& &
\label{chi-4}
(V \times S)^{1TR}_{aa;bb} \equiv
\frac{1}{2N_f}
[ V \times S ]^{\rm sd';s'd}_{aa;bb}
\ea
where the notations in the righthand side of the equations
are defined in Eqs. (\ref{note_1}-\ref{note_4}).
Using the lattice Ward identity Eq. (\ref{ward1}),
we obtain
\ba
& &(m_{d}+m_{s})(m_{d^{'}}+m_{s^{'}})
\sum_{t,t^{'}} {\cal M}_K^{1TR}(t,t^{'})
= \sum_{\mu}\sum_{(ab;ba),(aa;bb)}
\nn \\
& &
\langle 0 \mid
[
\bar{\chi}_{\rm s'}
\overline{ (\gamma_{\mu5} \otimes \xi_{5}) }
(\chi_{\rm s} \bar{\chi}_{\rm s}
- \chi_{\rm d} \bar{\chi}_{\rm d})
\overline{ (\gamma_{\mu5} \otimes \xi_{5}) }
\chi_{\rm s'}
\nn \\
& & \hspace{0.5 in}+
\bar{\chi}_{\rm s'}
\overline{ (\gamma_{\mu} \otimes I) }
(\chi_{\rm s} \bar{\chi}_{\rm s}
- \chi_{\rm d} \bar{\chi}_{\rm d})
\overline{ (\gamma_{\mu} \otimes I) }
\chi_{\rm s'}
\nn \\
& & \hspace{0.5 in}+
\bar{\chi}_{\rm d'}
\overline{(\gamma_{\mu5} \otimes \xi_{5}) }
(\chi_{\rm d} \bar{\chi}_{\rm d}
- \chi_{\rm s} \bar{\chi}_{\rm s})
\overline{(\gamma_{\mu5} \otimes \xi_{5}) }
\chi_{\rm d'}
\nn \\
& & \hspace{0.5 in}+
\bar{\chi}_{\rm d'}
\overline{(\gamma_{\mu} \otimes I) }
(\chi_{\rm d} \bar{\chi}_{\rm d}
- \chi_{\rm s} \bar{\chi}_{\rm s})
\overline{(\gamma_{\mu} \otimes I) }
\chi_{\rm d'}
\nn \\
& & \hspace{0.5 in} +
\bar{\chi}_{\rm s'}
\overline{ (\gamma_{\mu5} \otimes I) }
(\chi_{\rm s} \bar{\chi}_{\rm s}
- \chi_{\rm d} \bar{\chi}_{\rm d})
\overline{ (\gamma_{\mu5} \otimes I) }
\chi_{\rm s'}
\nn \\
& & \hspace{0.5 in}+
\bar{\chi}_{\rm s'}
\overline{ (\gamma_{\mu} \otimes \xi_{5}) }
(\chi_{\rm s} \bar{\chi}_{\rm s}
- \chi_{\rm d} \bar{\chi}_{\rm d})
\overline{ (\gamma_{\mu} \otimes \xi_{5}) }
\chi_{\rm s'}
\nn \\
& & \hspace{0.5 in}+
\bar{\chi}_{\rm d'}
\overline{(\gamma_{\mu5} \otimes I) }
(\chi_{\rm d} \bar{\chi}_{\rm d}
- \chi_{\rm s} \bar{\chi}_{\rm s})
\overline{(\gamma_{\mu5} \otimes I) }
\chi_{\rm d'}
\nn \\
\label{l_chiral2}
& & \hspace{0.5 in}+
\bar{\chi}_{\rm d'}
\overline{(\gamma_{\mu} \otimes \xi_5) }
(\chi_{\rm d} \bar{\chi}_{\rm d}
- \chi_{\rm s} \bar{\chi}_{\rm s})
\overline{(\gamma_{\mu} \otimes \xi_5) }
\chi_{\rm d'}
]
\mid 0 \rangle \ ,
\ea
%
%
where the details of the derivation are given
in appendix C.
As one can see in the above, Eq.
(\ref{l_chiral2}) is a lattice
version of Eq. (\ref{c_chiral3}).
The key point is that in the one spin trace formalism
a given choice of the vector and axial channel must be
augmented by a particular
choice of chiral partner channel
in order to have the
correct continuum chiral behavior.
Therefore, the implementation of
the continuum chiral behavior
in the one spin trace formalism forces us
to use at least eight operators rather than
four operators in the two spin trace formalism.
\subsection{Chiral Limit}
In this section,
we would like
to discuss the leading order and the next-to-leading
order chiral behavior of $ B_K $.
Let us consider the Fourier transform of the
weak matrix elements of $ B_K $
\ba
C^{1TR}_{KK}(E,E') & \equiv &
\sum_{t,t'} \exp(+iEt) M_K^{1TR}(t,t') \exp(-iE't')
\\
C^{2TR}_{KK}(E,E') & \equiv &
\sum_{t,t'} \exp(+iEt) M_K^{2TR}(t,t') \exp(-iE't') \ ,
\ea
where $ {\cal M}^{1TR}_K(t,t') $
and $ {\cal M}^{2TR}_K(t,t') $ are defined
in Eq. (\ref{m_1tr})
and Eq. (\ref{m_2tr}) respectively.

For the remainder of this discussion, let us drop the
label of the spin trace formalism as everything
applies to both cases.
Let us use the following parametrizations
\cite{sharpe1}:
\ba
	\ckk(E,E') &=& \frac{\sqrt{Z_{K }}}{{E }^2+m^2_{K }}
 		    \frac{\sqrt{Z_{K'}}}{{E'}^2+m^2_{K'}}
		    N_f \akk(E,E') + {\cal O}(a)  \\
	C_K (E)    &\equiv& (m_{s'}+m_{d'}) \ckk(E,0)
		  = \frac{\sqrt{Z_{K }}}{{E }^2+m^2_{K }}
		  \sqrt{2} N_f^{3/2} \ak (E ) + {\cal O}(a)  \\
	C_{K'}(E')   &\equiv& (m_{s }+m_{d }) \ckk(0,E')
	          = \frac{\sqrt{Z_{K'}}}{{E'}^2+m^2_{K'}}
		  \sqrt{2} N_f^{3/2} \akp(E') + {\cal O}(a)  \\
	C &\equiv& (m_s+m_d)(m_{s'}+m_{d'}) \ckk(0,0)
		    = 2 N_f^{2}   {\cal A} + {\cal O}(a)  \;.
\ea
In Ref. \cite{tool-kit},
it is shown that
\be
\sqrt{Z_K} = \sqrt{2N_f} f_K \frac{m_K^2}{m_s+m_d}
+ {\cal O}(m_K^4) \;,
\ee
which implies that $ m_K^2 \propto m_s+m_d $.
Consequently, $\akk$ has to be expanded
to $ \CO(m^n,E^{2n}) $
\ba
\label{exp-ee}
\akk(E,E') &=& \alpha_{00}
+ \alpha_{10} E + \alpha_{01} E'
+ \alpha_{20} E^2 + \alpha_{11} EE' + \alpha_{02} E'^2
+ \CO(E^3),
\ea
where the coefficients $ \alpha $
are functions of the quark masses
up to the appropriate power.
At this point we neglect
chiral logarithms and assume analyticity
of $ \akk $.

For local operators in the limit of
vanishing lattice spacing $ a \rightarrow 0 $,
one has
\ba
\label{time-r}
{\cal M}_K(t,t')= {\cal M}_K(-t,-t')
\ea
and since $ {\cal M}_K(t,t') $ is real
it follows that $\ckk $ is real as well.
{}From the definition of $ \ckk(E,E') $, we notice that
\ba
& & \mbox{Re} [\ckk(E,E')] = \mbox{function of even power of }
 E,E'\mbox{ product} \\
& & \mbox{Im} [\ckk(E,E')] = \mbox{function of odd power of }
 E,E'\mbox{ product}
\ea
This implies that only even powers remain
which confirms that $ \ckk(E,E') $ does not have a
linear term in either $E$ or $E'$
in the continuum limit as $ a \rightarrow 0 $
\cite{sharpe1}.

However, on a lattice of finite lattice spacing $a$,
the operator ($ {\cal O}^{Latt}_{2TR}$ or
$ {\cal O}^{Latt}_{1TR} $ is in general non-local in time.
In this case, Eq. (\ref{time-r}) does not hold
any more.
Instead, we have the following
\ba
{\cal M}_K(t,t^{'})
= {\cal M}_K(1 - t, 1 - t^{'})
\ea
which transfers the lattice into itself and
is an exact symmetry of the non-local
operator $ {\cal O}^{Latt} $.

Therefore, there exist terms with odd power in
either $E$ or $E'$ which  will be
of order $ a $. This implies that
in Eq. (\ref{exp-ee}),
$ \alpha_{10} $ and $ \alpha_{01} $ are of order
$ a $.
%
%
%
We finally have
\ba
\ckk(0,0) &=& \frac{\sqrt{Z_{K }}}{m^2_{K }}
	\frac{\sqrt{Z_{K'}}}{m^2_{K'}}
N_f \alpha_{00}  + \CO(a)
\\
&=& 2 N_f^2 \frac{f_K}{m_s+m_d}
\frac{f_{K'}}{m_{s'}+m_{d'}} \alpha_{00}
			+ \CO(a) \;,
\ea
which has been shown to vanish for $m_s=m_d$
and for $m_{s'}=m_{d'}$ in the previous sections.
Therefore, the quark mass independent term in
$\alpha_{00}$ must vanish and we obtain
\be
\alpha_{00} = \bar{\alpha}_{00}
	(m_s-m_d) (m_{s'}-m_{d'}) \;.
\ee
The $\CO(a)$ term should also vanish in the chiral limit.
In the physical case of $ K=K' $, we thus find
for the on-shell matrix element
$ {\cal A }_{KK}(-\i m_K,-\i m_K) $
\be
\label{exp-kk}
{\cal A }_{KK}^{\rm{on-shell}} =
\alpha_{00} -\g m_K^2 + \CO(m_K^4) + \CO(am_K) \;,
\ee
which is of the desired form.
Clearly, the
important condition to obtain the continuum $ B_K $
is that $ am_K \ll 1 $ and that the ratio $ m_s/m_d $
should be chosen properly
(at least $ m_s/m_d \gg 1 $)
in the numerical simulation in order to achieve a physical
value of $ \alpha_{00} $.
\section{Renormalization}
In this section, we will give a review of
continuum renormalization
\cite{buras,alti,collins}
and discuss the perturbative
corrections to the lattice
bilinear and four-fermion operators
in the staggered fermion formulation.
We adopt the powerful formulation of Ref. \cite{sheard0}.
This formulation of Ref. \cite{sheard0} was extended in
Refs. \cite{sheard1,sharpe3,Japan1,sharpe4}
to include matching all kinds of
complicated lattice operators
to the continuum operators at the one loop level.
The renormalization and matching of
lattice bilinear operators
was studied
in Refs. \cite{sheard0,sharpe3,Japan1}.
The renormalization of four-fermion operators
in the two spin trace formalism
and their connection to the continuum operators
was investigated in
Refs. \cite{sheard1,Japan1,sharpe4}.

In addition to the above, we will also
explain the renormalization
of four fermion operators
in the one spin trace formalism,
including the renormalization of chiral partner
operators.
\subsection{Feynman rules}
Let us briefly give the relevant Feynman rules
for the staggered fermion action and lattice composite
operators.
In the general covariant gauge
with a gauge parameter $ \alpha $
(for example, $ \alpha $ = 1 and 0 corresponding
to Feynman and
Landau gauge respectively), the gluon propagator is
\ba
D_{\mu \nu}^{IJ}(k) =
\frac{ \delta_{IJ}\delta_{\mu\nu} }
{ \sum_{\beta} \frac{4}{a^2} \sin^{2}( \frac{1}{2} ak_{\beta} ) }
-
(1-\alpha)
\frac{ \delta_{IJ} \frac{4}{a^2}
\sin( \frac{1}{2} ak_{\mu} ) \sin( \frac{1}{2} ak_{\nu} )}
{ \sum_{\beta} \frac{4}{a^2} \sin^{2}( \frac{1}{2} ak_{\beta} ) } \ ,
\ea
where $ I $ and $ J $ represent the color indices in the
adjoint representation of SU(3).

The fermion propagator is
\ba
S(p,-q)_{\mbox{\small ab}} & = & \delta_{\mbox{\small ab}}
\frac{ m \bar{\delta}(p-q) +
\frac{i}{a} \sum_{\mu} \sin(ap_{\mu})
\bar{\delta}(p-q+\pi_{\eta_{\mu}}/a )  }
{ \sum_{\beta} \frac{1}{a^2} \sin^2( ap_{\beta} ) + m^2 }
\\
& = & \delta_{\mbox{\small ab}}
\frac{ m \bar{\delta}(p-q) +
\frac{i}{a} \sum_{\mu} \sin(aq_{\mu})
\bar{\delta}(p-q+\pi_{\eta_{\mu}}/a )  }
{ \sum_{\beta} \frac{1}{a^2} \sin^2( aq_{\beta} ) + m^2 }
\ea
where
\ba
\pi_{\eta_{\mu}} & = & \pi \cdot
\sum_{\nu = 1}^{\mu-1} \hat{\nu} \ , \ \
\hat{\nu} = \mbox{unit vector along $ \nu $ direction}
\\
\mbox{i.e. } \
\pi_{\eta_{1}} & = & (0,0,0,0),\
\pi_{\eta_{2}} = (\pi,0,0,0),\
\pi_{\eta_{3}} = (\pi,\pi,0,0),\
\pi_{\eta_{4}} = (\pi,\pi,\pi,0),
\nonumber
\\
\bar{\delta}(p) & = &
(2\pi)^4 \sum_{n} \delta(p + \frac{2\pi}{a}n)
\  \mbox{ with } \
n \equiv (n_1,n_2,n_3,n_4) \in I^{4}
\ea
and a and b are color indices in the fundamental
representation of SU(3), $a$ is the lattice spacing
and $ -\pi/a < p,q,k \leq \pi/a $.

The one-gluon and two-gluon vertices associated with
the staggered fermion action are
\ba
V_{\mu}^{I}(p, -q, k)_{\mbox{\small ab}} & = &
-ig T^{I}_{\mbox{\small ab}} \cos( a [\frac{k_{\mu}}{2} +p_{\mu}] )
\bar{\delta}( p-q+k+\pi_{\eta_{\mu}}/a )
\nonumber
\\
& = &
-ig T^{I}_{\mbox{\small ab}} \cos( a [\frac{k_{\mu}}{2}-q_{\mu}] )
\bar{\delta}( p-q+k+\pi_{\eta_{\mu}}/a )
\\
V_{\mu\nu}^{IJ}(p,-q; k_1,k_2)_{\mbox{\small ab}} & = &
-ia g^2 \delta_{\mu\nu}\frac{1}{2}
\{T^{I},T^{J}\}_{\mbox{\small ab}}
\sin( a[p_{\mu} +\frac{ k_{1\mu} + k_{2\mu} }{2}])
\nonumber
\\
& & \bar{\delta}(p-q+k_1+k_2+\pi_{ \eta_{\mu} }/a)\ ,
\ea
where $ p $ and $ q $
are the fermion momenta
and $ k, k_1, k_2 $
are out-going gluon momenta.
Since products of gauge links are present in the definition
of the gauge invariant operators in Eq.(\ref{bilinear}),
the vertices of an operator will generally have
external gluon lines.
The operator vertices
are
\ba
\label{zerogluon}
M_{SF}^{(0)}(p,-q)_{\mbox{\small ab}} & = &
\delta_{\mbox{\small ab}}\frac{1}{N_f^2}
\sum_{AB} e^{iap \cdot A}
( \overline{ \gamma_S \otimes \xi_F} )_{AB}
e^{-iaq \cdot B},
\\
M_{SF;\mu}^{(1)I}(p,-q;k)_{\mbox{\small ab}} & = &
iag T^{I}_{\mbox{\small ab}}
\frac{1}{N_f^2}
\sum_{AB} e^{iap \cdot A}
( \overline{ \gamma_S \otimes \xi_F} )_{AB}
e^{-iaq \cdot B}
\nonumber
\\
\label{onegluon}
& & \cdot
(B_{\mu}-A_{\mu})
f^{\mu}_{(AB)}(ak),
\\
M_{SF;\mu\nu}^{(2)IJ}(p,-q;k_1,k_2)_{\mbox{\small ab}}
& = &
\frac{1}{2} (iag)^2 \frac{1}{2}
\{T^{I},T^{J}\}_{\mbox{\small ab}}
\frac{1}{N_f^2}
\sum_{AB} e^{iap \cdot A}
( \overline{ \gamma_S \otimes \xi_F } )_{AB}
e^{-iaq \cdot B }
\nonumber
\\
\label{twogluon}
& & \cdot
(B_{\mu}-A_{\mu})
(B_{\nu}-A_{\nu})
g^{\mu\nu}_{(AB)}(ak_1, ak_2) \ ,
\ea
where the superscript in curved brackets
$ (i) $, $ i = 0,1,2,\cdots $
denotes the number of
emitted gluons.
The function $ f^{\mu}_{(AB)}(\phi \equiv ak) $ is
\ba
f^{\mu}_{(AB)}(\phi) & \equiv & \frac{1}{12}
e^{i \phi \cdot A} e^{i \frac{\phi}{2} \cdot \Delta_{\mu} }
\nonumber
\\
& &
\sum_{ \nu \neq \mu} [ 1 +
e^{i(\sum_{\rho} \phi \cdot \Delta_{\rho})
-i \phi \cdot \Delta_{\mu} }
+ e^{i(\sum_{\rho} \phi \cdot \Delta_{\rho})
-i \phi \cdot \Delta_{\mu} -i \phi \cdot \Delta_{\nu} }
+ e^{i \phi \cdot \Delta_{\nu} } ]
\nonumber
\\
& = &
\frac{1}{12} e^{i \phi \cdot A}
\sum_{\nu \neq \mu } \sum_{ j = 1}^{4}
e^{i(B-A) \cdot \theta^{(j)}_{\mu\nu}( \phi) }
\ea
where
\ba
\phi & = & \sum_{\rho} \phi_{\rho} \hat{\rho}\ , \\
\theta^{(1)}_{\mu\nu}(\phi) & = &
\frac{1}{2} \phi_{\mu} \hat{\mu},
\\
\theta^{(2)}_{\mu\nu}(\phi) & = &
\frac{1}{2} \phi_{\mu} \hat{\mu}
+ \phi_{\nu} \hat{\nu},
\\
\theta^{(3)}_{\mu\nu}(\phi) & = &
\phi
-\frac{1}{2} \phi_{\mu} \hat{\mu},
\\
\theta^{(4)}_{\mu\nu}(\phi) & = &
\phi
-\frac{1}{2} \phi_{\mu} \hat{\mu}
-\phi_{\nu} \hat{\nu} \ .
\ea
and $ \Delta_{\mu} = (B_{\mu}-A_{\mu}) \hat{\mu} $
is defined in Eq. (\ref{delta}).
All that is required to know about $ g^{\mu\nu}_{(AB)} $
at the one loop level is the following tadpole contribution
\ba
g^{\mu\nu}_{(AB)}(\phi, -\phi) & = & 1 \ \mbox{for $ \mu=\nu $ }
\\
& = & \frac{1}{4!}
e^{i\frac{ \phi }{2} \cdot (\Delta_{\mu} + \Delta_{\nu})}
[6 + 2 \sum_{\rho \neq \mu,\nu}
e^{i \phi \cdot \Delta_{\rho} }
+ 2 e^{i \phi \cdot \sum_{\rho \neq \mu,\nu} \Delta_{\rho} }]
+ \mbox{h.c.   for $ \mu \neq \nu $ } \ \ .
\ea
In order to carry out a perturbative expansion of the
Landau gauge operators, we need only the Feynman rule
in Eq.(\ref{zerogluon}), because all the gauge
links $ {\cal U} $ are replaced by the identity
\cite{sharpe1,sharpe3,sharpe4}.
The above Feynman rules given in analytic form
are also shown graphically in Figure 1.
\footnote{
The Feynman rules given in this section
are consistent with those in Ref.
\cite{sheard0,sharpe3,Japan1}
except for a few minor typing mistakes.}

Feynman rules for the four fermion operators can be
easily obtained
from products of two bilinear operators as long as
the color and hypercube indices are arranged carefully.
They will not be given here.
\subsection{Bilinear Operators}
Let us review one-loop renormalization
and matching of the bilinear operators.
Consider the bare Green's function
\ba
\langle q(x) \bar{q}(y) {\cal O}_{SF} \rangle \ ,
\ea
where $ {\cal O}_{SF} $ is a bilinear operator with
spin $ S $ and with flavor $ F $.
There are 3 one loop Feynman diagrams \cite{buras,alti}
in the continuum.
We can write upto one loop
\ba
\label{bilinearc}
{\cal O}^{Cont}_{SF} = \sum_{S'F'}
( \delta_{SS'} \delta_{FF'} +
\frac{g^2}{(4\pi)^2} Z^{Cont}_{SF;S'F'} )
{\cal O}^{Cont(0)}_{S'F'} \ ,
\ea
where the superscript $ (i) $,
$ i = 0,1,2,\cdots $
represents the number of loops and
\ba
Z^{Cont}_{SF;S'F'} \equiv
\delta_{SS'}\delta_{FF'} \Gamma_S \log(\frac{\mu}{\kappa})
+ \delta_{SS'}\delta_{FF'} C^{Cont}_{S}
+ \delta_{SS'}\delta_{FF'} R_{S} \ .
\ea
Here $ \mu $ is the renormalization scale
and $ \kappa $ is the gluon mass for the
infra-red regulator.
$ \Gamma_S $ is the anomalous dimension matrix
as follows
\ba
\label{anomalous1}
\Gamma_S & = & \frac{8}{3}(\sigma_S -1)
\\
\sigma_S \gamma_S & = & \frac{1}{4}\sum_{\mu,\nu}
\gamma_{\mu}\gamma_{\nu}\gamma_{S}
\gamma_{\nu} \gamma_{\mu}\ .
\ea
Note that the vector and axial currents
have vanishing anomalous dimensions.
$ C^{Cont}_{S} $ is the finite constant term which
depends on
the regularization and renormalization scheme.
For the $ \overline{MS} $ scheme with NDR
(naive dimensional regularization) \cite{buras},
\ba
\sigma_S & = & (4,1,0,1,4) \ , \\
C^{Cont}_{S} & = &
(\frac{10}{3}, 0 ,\frac{2}{3}, 0, \frac{10}{3})
\mbox{  for  } \gamma_S = (I,\gamma_{\mu},
\sigma_{\mu\nu},\gamma_{\mu5},\gamma_{5} )
\mbox{ respectively.}
\ea
$ R_{S} $ refers to those universal terms
which contain all the external momentum dependence
and are independent of the regularization scheme.

On the lattice, there are
eight Feynman diagrams (Figure 2 (a)-(h))
contributing to the bilinear operator.
Only half of the value of the
wave-function renormalization  component
self-energy diagrams
(Figure 2 (e), (f), (g) and (h))
contribute to the lattice operator
since the other half are absorbed
in the wave-function renormalization
of the external quark fields.
On the lattice,
we have instead of Eq. (\ref{bilinearc})
\ba
\label{bilinearl}
{\cal O}^{Latt}_{SF}
= \sum_{S'F'}
(\delta_{SS'} \delta_{FF'} +
\frac{g^2}{(4\pi)^2} Z^{Latt}_{SF;S'F'} )
{\cal O}^{Latt(0)}_{S'F'} \ ,
\ea
where
\ba
Z^{Latt}_{SF;S'F'} \equiv
-\delta_{SS'}\delta_{FF'} \Gamma_S \log(a{\kappa})
+ C^{Latt}_{SF;S'F'}
+ \delta_{SS'}\delta_{FF'} R_{S} \ .
\ea

On the lattice, $ \Gamma_S $ is
the same as in the continuum
(see Eq. (\ref{anomalous1}))
since the anomalous dimension at the one loop level
is universal, i.e. independent
of the regularization and renormalization schemes
\cite{collins}.
$ C^{Latt}_{SF;S'F'} $ is
the finite constant term unique to
the lattice regularization scheme.
For the Landau gauge definition
of the axial current
($ \gamma_S = \gamma_{\mu5} $ and $ \xi_{F} = \xi_{5} $)
the finite term has been calculated
in Ref. \cite{sharpe3,Japan1}
\ba
C^{Latt}_{\mu5,5;\mu5,5} = 12.232 \ .
\ea
$ C^{Latt}_{SF;S'F'} $ for bilinear operators
with various spin-flavor structures can be found in
Ref. \cite{sharpe3,Japan1}.
$ R_S $ is the same as that in the continuum.

In the classical limit (i.e. at tree level
as the lattice spacing $ a $ goes to zero),
the lattice operator
can be related to the continuum operator as follows
\ba
\label{connect}
{\cal O}^{Latt(0)}_{SF} = {\cal O}^{Cont(0)}_{SF} +
{\cal O}(a) \ .
\ea
In a perturbative calculation on the lattice,
terms of order $ a $ or higher are supposed
to be negligible
as $ a \rightarrow 0 $.
At one loop (and higher) order the lattice and the
continuum operators will differ and a carefully
constructed mixture of lattice operators is needed
to reproduce the desired continuum operator.
{}From Eqs.(\ref{bilinearc}),
(\ref{bilinearl}), (\ref{connect}), we can connect
the lattice operator with the continuum operator
through one loop via
\ba
{\cal O}^{Cont}_{SF} & = &
\sum_{S'F'} \left[ \delta_{SS'}\delta_{FF'} +
\frac{ g^2 }{ (4\pi)^2 }
\left( Z^{Cont}_{SF;S'F'} - Z^{Latt}_{SF;S'F'} \right)
\right] {\cal O}^{Latt}_{S'F'}
\nonumber \\
& = & \sum_{S'F'}
\left[ \delta_{SS'}\delta_{FF'}
+ \frac{ g^2 }{ (4\pi)^2 }
\delta_{SS'} \delta_{FF'} \ln(\mu a) +
\right.
\nonumber \\
& & \left. \frac{ g^2 }{ (4\pi)^2 }
\left(
\delta_{SS'} \delta_{FF'} C^{Cont}_S
-C^{Latt}_{SF;S'F'}
\right) \right]
{\cal O}^{Latt}_{S'F'}
\ea

In order to understand this operator mixing better,
let us consider some important
symmetry properties of the bilinear
operators \cite{verstegen,sheard0,sheard1,sharpe4}.
The {\em distance} of the lattice bilinear
operator with spin $ S $ and flavor $ F $
\cite{sharpe1,sharpe4} is defined as
\ba
\Delta = \sum_{\mu = 1}^{4}
\mid  S_{\mu} - F_{\mu}  \mid^2 \ ,
\ea
and corresponds
to the number of links between the
quark and anti-quark fields.
The {\em distance parity}
of the lattice bilinear operators
is given by $ (-1)^{\Delta} $.

The massless staggered fermion action
has a $ U_A(1) $ symmetry:
\ba
\label{transf-1}
\chi & \longrightarrow & \exp[i\alpha
( \overline{ \gamma_5 \otimes \xi_5} )]\chi \
\\
\bar{\chi} & \longrightarrow &  \bar{\chi}
\exp[i\alpha
( \overline{ \gamma_5 \otimes \xi_5} )]
\ea
with
\ba
\exp[i \alpha (\overline{ \gamma_5 \otimes \xi_5 } )]
= \cos(\alpha)(\overline{ I \otimes I }) +
i\sin(\alpha) (\overline{ \gamma_5 \otimes \xi_5})
\ea
Under this transformation,
bilinear operators transform as:
\ba
\bar{\chi} (\overline{ \gamma_S \otimes \xi_F }) \chi
& \longrightarrow &
[ \cos^2(\alpha) - (-1)^{\Delta}\sin^2(\alpha)]
\bar{\chi} (\overline{ \gamma_S \otimes \xi_F }) \chi
\nonumber
\\
& &
+ i \sin(\alpha) \cos(\alpha)[1 + (-1)^{\Delta}]
\bar{\chi} (\overline{ \gamma_{S5} \otimes \xi_{F5} }) \chi.
\ea
Therefore
even-distance bilinear operators
$ \bar{\chi} (\overline{ \gamma_{S} \otimes \xi_{F} }) \chi $
(i.e. $ \Delta $ is even)
can be rotated into
\ba
\bar{\chi} (\overline{ \gamma_{S5} \otimes \xi_{F5} }) \chi
\ ,
\nonumber
\ea
while odd-distance bilinear operators are invariant.
This guarantees that even and odd distance bilinear
operators never mix with each other,
since they belong to different
irreducible representations respectively
with respect to this $ U_A(1) $ axial
rotation group. Furthermore,
it follows that the two even-distance bilinear operators
$ \bar{\chi} (\overline{ \gamma_{S} \otimes \xi_{F} }) \chi $
and
$ \bar{\chi} (\overline{ \gamma_{S5} \otimes \xi_{F5} }) \chi $
are  renormalized identically (i.e. have
the same anomalous dimension and the same finite term
to all orders in the perturbative expansion).

We will now prove that the same is true
for odd-distance bilinear operators
from $ U_A(1) $ symmetry.
This is in contrast to Sharpe and Patel,
who argued in Ref. \cite{sharpe4} that
the result does {\em not} follow
from $ U_A(1) $ symmetry.

Introducing a separate staggered fermion
for each continuum quark flavor,
we have $ \chi_u $, $ \chi_d $ and $ \chi_s $
for three flavor QCD.
Besides, for each continuum quark there
are four degenerate flavors coming from
the staggered fermion action.
We then introduce a separate $ U_A(1) $
transformation for each continuum quark flavor
\ba
\label{transf-2}
\chi_{\rm u} \longrightarrow \exp[i\alpha
( \overline{ \gamma_5 \otimes \xi_5} )] \chi_{\rm u} &,&
\bar{\chi}_{\rm u} \longrightarrow \bar{\chi}_{\rm u} \exp[i\alpha
( \overline{ \gamma_5 \otimes \xi_5} )]
\\
\label{transf-3}
\chi_{\rm d} \longrightarrow \exp[i\beta
( \overline{ \gamma_5 \otimes \xi_5} )] \chi_{\rm d} &,&
\bar{\chi}_{\rm d} \longrightarrow \bar{\chi}_{\rm d} \exp[i\beta
( \overline{ \gamma_5 \otimes \xi_5} )]
\\
\label{transf-4}
\chi_{\rm s} \longrightarrow \exp[i\eta
( \overline{ \gamma_5 \otimes \xi_5} )] \chi_{\rm s} &,&
\bar{\chi}_{\rm s} \longrightarrow \bar{\chi}_{\rm s} \exp[i\eta
( \overline{ \gamma_5 \otimes \xi_5} )] \ .
\ea
Under this set of three $ U_A(1) $ transformations,
the bilinear operators are transformed into
\ba
\bar{\chi}_{\rm s} (\overline{ \gamma_S \otimes \xi_F })
\chi_{\rm d}
& \longrightarrow &
\left[ \cos(\beta)\cos(\eta)-
(-1)^{\Delta}\sin(\beta)sin(\eta) \right]
\bar{\chi}_{\rm s} (\overline{ \gamma_S \otimes \xi_F })
\chi_{\rm d}
\nonumber
\\
& &
+ i
\left[ \cos(\eta) \sin(\beta)
+ (-1)^{\Delta} \sin(\eta) \cos(\beta) \right]
\bar{\chi}_{\rm s} (\overline{ \gamma_{S5} \otimes \xi_{F5} })
\chi_{\rm d}
\ea
Even and odd distance bilinear operators still belong to
different irreducible representations.
However, now both even
and odd distance bilinear operators
can be rotated into
\ba
\bar{\chi}_{\rm s} (\overline{ \gamma_{S5} \otimes \xi_{F5} })
\chi_{\rm d}
\ .
\nonumber
\ea
This insures that
for any spin-flavor structure,
the two bilinear operators
$ \bar{\chi}
(\overline{ \gamma_{S} \otimes \xi_{F} }) \chi $
and
$ \bar{\chi}
(\overline{ \gamma_{S5} \otimes \xi_{F5} }) \chi $
are  renormalized identically
regardless of their distance.
\subsection{Four-Fermion Operators}
We now discuss the renormalization of four-fermion
operators which are of the general form
\ba
[\bar{\chi}_{f_1}
(\overline{ \gamma_S \otimes \xi_F} )
\chi_{f_2}]
[\bar{\chi}_{f_3}
(\overline{ \gamma_{S'} \otimes \xi_{F'} } )
\chi_{f_4}] \ ,
\ea
where the spin-flavor
structure $ SF;S'F'$ is arbitrary,
allowing for $ 256^2 = 65536 $
possible combinations.
But we are interested only
in those operators which
contribute to the phenomenological
weak matrix elements
(especially $ B_K $).
These operators
are the so-called {\em diagonal operators}
\cite{sharpe1,Japan1,sharpe4}
with $ S = S'$ and $ F = F' $ and
belong to the representation
$ I $ of the Euclidean lattice rotation group
\cite{verstegen,sheard1,sharpe1,sharpe4}.
For the $ B_K $ calculation,
we can restrict the set of operators further
to the following bilinear structures
\ba
\label{bilinear4}
(\overline{ \gamma_S \otimes \xi_F} ) \in
\left\{
(\overline{ \gamma_{\mu} \otimes I} ),
(\overline{ \gamma_{\mu} \otimes \xi_{5}} ),
(\overline{ \gamma_{\mu5} \otimes I} ),
(\overline{ \gamma_{\mu5} \otimes \xi_{5}} ).
\right\}
\ea
The bilinear operators corresponding to
Eq. (\ref{bilinear4})
belong to $ J^{PC} = (\frac{1}{2},\frac{1}{2})^{+-} $,
$ \overline{(\frac{1}{2},\frac{1}{2})}^{+-} $,
$ (\frac{1}{2},\frac{1}{2})^{--} $ and
$ \overline{(\frac{1}{2},\frac{1}{2})}^{--} $
respectively \cite{verstegen,mandula,sheard0,sheard1}.
Any linear combination of the diagonal four-fermion
operators with the spin-flavor structure $SF;SF$ given in
Eq. (\ref{bilinear4})
transforms as $ J^{PC} = I^{++} $
\cite{verstegen,mandula,sheard1}.
\subsubsection{Continuum Four-Fermion Operators}
Let us consider the following Green's function in the continuum
\ba
\langle q(x_1) q(y_1) \bar{q}(x_2) \bar{q}(y_2)
{\cal O}^{Cont}_{S;S} \rangle \ ,
\ea
where $ {\cal O}^{Cont}_{S;S} $
is a four-fermion operator
with the spin structure $ S;S $.
{}From Eq. (\ref{deltas2}),
the continuum four-fermion operator for $ B_K $
($ \Delta S = 2 $) is
\ba
O^{Cont}_{\Delta S = 2} =
\bar{s}^{a} \gamma_{\mu} (1-\gamma_{5}) d^{a}
\bar{s}^{b} \gamma_{\mu} (1-\gamma_{5}) d^{b} \ .
\ea
There are 10 Feynman diagrams \cite{buras,alti}
contributing
to the continuum four-fermion operator
at the one loop level.
The continuum $ \Delta S = 2 $ operator
$ {\cal O}^{Cont}_{\Delta S = 2} $
\cite{buras,alti,wise2}
can be written upto one loop as
\ba
\label{fourfc}
{\cal O}^{Cont}_{\Delta S = 2} & = & \left[
1 + \frac{g^2}{(4\pi)^2} Z^{Cont}_{\Delta S=2} \right]
{\cal O}^{Cont(0)}_{\Delta S=2} \ ,
\ea
where the superscript $(i)$,  $ i = 0,1,2,\cdots $
represents the number of loops and
\ba
\label{zcont}
Z^{Cont}_{\Delta S=2} \equiv
\Gamma_{\Delta S=2} \log(\frac{\mu}{\kappa})
+ C^{Cont}_{\Delta S=2}
+ R_{\Delta S=2}.
\ea
$ \Gamma_{\Delta S=2} $ is the anomalous dimension of the
operator $ {\cal O}^{Cont}_{\Delta S=2} $.
We obtain \cite{wise2}
\ba
\Gamma_{\Delta S=2} = -4 \ .
\ea
$ C^{Cont}_{\Delta S=2} $ is a finite constant term which
depends on the regularization and renormalization scheme.
For the $ \overline{MS} $ scheme
with NDR \cite{buras,Japan1},
\ba
C^{Cont}_{\Delta S=2} = \frac{11}{12}
\Gamma_{\Delta S=2} \; .
\ea
$ R_{\Delta S=2} $ refers to those universal finite
terms which contain
all the external momentum dependence, independent
of the regularization and renormalization scheme.
\subsubsection{Lattice Two Spin Trace Operators}
On the lattice the Green's function is written as
\ba
\langle \chi(x_1) \chi(y_1) \bar{\chi}(x_2) \bar{\chi}(y_2)
{\cal O}^{Latt}_{SF;SF,2TR} \rangle \ ,
\ea
where $ {\cal O}^{Latt}_{SF;SF,2TR} $
belongs to the
two spin trace formulation
defined in section 3.1.
Since the vector and axial channels
in the two spin trace formalism
satisfy the continuum
chiral behavior separately, we need
to find
some linear combination of two spin trace operators
which obey the same normalization conditions that
specify the continuum operator
$ {\cal O}^{Cont}_{\Delta S = 2} $.
This is crucial so that $ R_{\Delta S=2} $
in Eq.(\ref{zcont})
cancels that of the lattice operator
when we make a connection between
the continuum and the lattice operators.

The lattice four-fermion operators for $ B_K $
in the two spin trace formalism are
\ba
& &
\frac{1}{N_f^2}
[\bar{\chi}^{a}_{\rm s}
(\overline{ \gamma_{\mu} \otimes \xi_5 } ) \chi^{b}_{\rm d}]
[\bar{\chi}^{b}_{\rm s'}
(\overline{ \gamma_{\mu} \otimes \xi_5 } ) \chi^{a}_{\rm d'}]
\Longleftrightarrow (V \times P)^{2TR}_{ab;ba}
\\
& &
\frac{1}{N_f^2}
[\bar{\chi}^{a}_{\rm s}
(\overline{ \gamma_{\mu} \otimes \xi_5 } ) \chi^{a}_{\rm d}]
[\bar{\chi}^{b}_{\rm s'}
(\overline{ \gamma_{\mu} \otimes \xi_5 } ) \chi^{b}_{\rm d'}]
\Longleftrightarrow (V \times P)^{2TR}_{aa;bb}
\\
& &
\frac{1}{N_f^2}
[\bar{\chi}^{a}_{\rm s}
(\overline{ \gamma_{\mu5} \otimes \xi_5 } ) \chi^{b}_{\rm d}]
[\bar{\chi}^{b}_{\rm s'}
(\overline{ \gamma_{\mu5} \otimes \xi_5 } ) \chi^{a}_{\rm d'}]
\Longleftrightarrow (A \times P)^{2TR}_{ab;ba}
\\
& &
\frac{1}{N_f^2}
[\bar{\chi}^{a}_{\rm s}
(\overline{ \gamma_{\mu5} \otimes \xi_5 } ) \chi^{a}_{\rm d}]
[\bar{\chi}^{b}_{\rm s'}
(\overline{ \gamma_{\mu5} \otimes \xi_5 } ) \chi^{b}_{\rm d'}]
\Longleftrightarrow (A \times P)^{2TR}_{aa;bb} \ ,
\ea
as given in Eq. (\ref{op-2tr}--\ref{op-2tr-4}).
They all belong to the $ I^{++} $ representation of
the Euclidean rotation group with discrete
parity and charge conjugation symmetry \cite{verstegen}.
There are a large number of
four-fermion operators in the same
$ I^{++} $ representation, which will mix with the
the above four-fermion operators.
Furthermore, on the lattice color one-trace operators
mix with color two-trace operators
as in the continuum.

Hence the general form
for the one-loop renormalization
of the four-fermion operators on the lattice is
\ba
\label{lattice-one-loop}
{\cal O}^{Latt(1)}_{i,SF;SF} =
\left[ \delta_{ij}\delta_{SS'} \delta_{SS''}
\delta_{FF'}\delta_{F'F''}
+ \frac{g^2}{(4\pi)^2}
Z^{Latt}_{ij,SF;S'F';S''F''} \right]
{\cal O}^{Latt(0)}_{j,S'F';S''F''} \ .
\ea
where $ {\cal O}^{Latt}_{i,SF;SF} $
stands for either spin trace formulation.
The subscript j of
$ {\cal O}^{Latt,(i)}_{j,SF;SF} $
is the number of color traces
and
\ba
Z^{Latt}_{ij,SF;S'F';S''F''} =
\delta_{S'S''} \delta_{FF'} \delta_{F'F''}
\left[- \Gamma_{ij,SS'} \ln(a\kappa) + R^{Latt}_{ij,SS'} \right]
+ C^{Latt}_{ij,SF;S'F';S''F''} \ .
\label{z-2tr}
\ea
$ \Gamma_{ij,SS'} $ is
the anomalous dimension matrix of the
operator $ {\cal O}^{Latt,(i)}_{j,SF;SF} $.
$ C^{Latt}_{ij,SF;S'F';S''F''} $ represents the
finite constant terms
which depend on the regularization
and renormalization scheme.
These terms generally cause the diagonal operator
to mix with operators of
non-diagonal flavor structure at the one loop level.
For $ B_K $, the total contribution of the
$ C^{Latt}_{ij,SF;S'F';S''F''} $ mixing
still belongs to the $ I^{++} $ representation.
$ R^{Latt}_{ij,SS'} $
represents those finite terms
which contain all
the external momentum dependence,
independent of the regularization
and renormalization scheme
as long as the anomalous
dimension of the lattice operator
is identical to that of
the continuum operator
as  $ a \rightarrow 0 $ .

We now have to show that
there exists
a particular linear combination of lattice operators
which has the same anomalous dimension as
the continuum operators at the one loop level.

There are altogether 50 Feynman diagrams
(25 diagrams each in Figure 3 and Figure 4)
contributing to the one-loop
radiative corrections to
$ {\cal O}^{Latt}_{SF;SF,2TR} $.
The diagrams in Figure 3
are for color two-trace operators
whereas the diagrams in Figure 4 are
for color one-trace
operators.
For Landau gauge operators,
only 28 out of the 50 diagrams
contribute to the one loop radiative corrections
and
20 out of these 28 diagrams
overlap with those of the bilinear operators.

Therefore, for Landau gauge operators
we only need to calculate 8
diagrams in addition to
those diagrams for the bilinear operator:
4 diagrams from color two-trace operators are
(g1), (g2), (g3) and (g4) in Figure 3 and 4
diagrams from color one-trace operators are
(g1), (g2), (g3) and (g4) in Figure 4.
The analytic results of the 28 diagrams for the Landau
gauge operators are explained
in appendix D and Ref. \cite{Japan1,sharpe4}.

For $ B_K $ in the two spin trace formalism,
using the scheme of Kilcup \& Sharpe
\cite{sharpe0,sharpe1,tool-kit},
we choose a minimum of  four operators
which mix with one another
through the anomalous dimension matrix at the one loop level
in order to extract the single operator that
appears in the physical continuum 3-flavor theory
$ {\cal O}^{Cont}_{\Delta S = 2} $.
Defining
\ba
\vec{\cal{O}}_{B_K,2TR}
\equiv \left( \begin{array}{c}
(V  \times P)^{2TR}_{ab;ba} \\
\vspace*{0.01 in} \\
(V  \times P)^{2TR}_{aa;bb} \\
\vspace*{0.01 in} \\
(A  \times P)^{2TR}_{ab;ba} \\
\vspace*{0.01 in} \\
(A  \times P)^{2TR}_{aa;bb}
\end{array} \right) \ ,
\ea
its anomalous dimension matrix is
\ba
\label{gamma}
\Gamma_{B_K} =
\left(
\begin{array}{cccc}
9  & -3 & -7 & -3 \\
0  & 0  & -6 & 2 \\
-7 & -3 & 9  & -3 \\
-6 & 2  &  0 &  0
\end{array}
\right) \  .
\ea
The eigenvalues of $ \Gamma_{B_K} $
are ($ -4 $, $-2 $, $ 8 $, $ 16 $) and the
corresponding eigen-operators are respectively
\ba
{\cal O}_{\Gamma = -4} & = &
(V \times P)^{2TR}_{ab;ba}
+ (V \times P)^{2TR}_{aa;bb}
+ (A \times P)^{2TR}_{ab;ba}
+ (A \times P)^{2TR}_{aa;bb}
\\
{\cal O}_{\Gamma = -2} & = &
-(V \times P)^{2TR}_{aa;bb}
+ (A \times P)^{2TR}_{aa;bb}
\\
{\cal O}_{\Gamma = 8} & = &
- (V \times P)^{2TR}_{ab;ba}
+ (V \times P)^{2TR}_{aa;bb}
- (A \times P)^{2TR}_{ab;ba}
+ (A \times P)^{2TR}_{aa;bb}
\\
{\cal O}_{\Gamma = 16} & = &
-3 (V \times P)^{2TR}_{ab;ba}
- (V \times P)^{2TR}_{aa;bb}
+ 3 (A \times P)^{2TR}_{ab;ba}
+ (A \times P)^{2TR}_{aa;bb} \ .
\ea
In the above,
only one eigen-operator
$ {\cal O}_{\Gamma = -4} $
has the same leading
logarithmic behavior as
the continuum $ \Delta S = 2 $
operator upto all orders in a perturbative expansion.

Hence by choosing $ {\cal O}_{\Gamma = -4} $ as the
lattice version of the continuum operator,
we can implement continuum leading logarithmic behavior.
It is also  guaranteed
that the total contribution of the
momentum-dependent finite terms
from $ R^{Latt,1TR}_{ij,SS'} $
will cancel that of $ R_{\Delta S = 2} $ when we make a
connection between the lattice
and the continuum operators.

For $ B_K $ in the two spin trace formalism,
the scheme-dependent finite-term matrix
$ C^{Latt}_{B_K,2TR} $ mixes
$ \vec{\cal{O}}_{B_K,2TR} $
with a large number of other operators in
the $ I^{++} $ representation.
In the two spin trace formalism,
the extra operators all
have different flavor structures,
and so we expect them to be suppressed
significantly in matrix elements
with the external eigenstates $ K^{0} $ and $ \bar{K}^{0} $
which have flavor $ \xi_5 $.
The scheme dependent finite terms
for Landau gauge operators $ \vec{\cal{O}}_{B_K,2TR} $
are explicitly
\ba
C^{Latt}_{B_K,2TR} =
\left(
\begin{array}{cccc}
37.446	& -2.9136 & -5.25285 & -2.2512 \\
0	& 28.706 & -4.5024 & 1.5008 \\
-5.25285& -2.2512 & 37.976  & -4.5043 \\
-4.5024 & 1.5008  &  0	   & 24.464
\end{array}
\right) \ .
\ea
We can match the lattice operators for $ B_K $
with the continuum $ \Delta S = 2 $ operator
by projecting the eigen-operator
$ {\cal O}_{\Gamma = -4} $ out as follows
\ba
& & {\cal O}^{(0)}_{\Gamma = -4} =
\vec{P}_{\Gamma = -4} \cdot \vec{\cal O}^{(0)}_{B_K,2TR}
= {\cal O}^{Cont(0)}_{\Delta S = 2} +
\mbox{ terms of order $ a $,}
\nonumber
\\
& &{\cal O}^{Cont(1)}_{\Delta S = 2}
= \vec{P}_{\Gamma = -4} \cdot
\left[ I + \frac{g^2}{(4\pi)^2}
\left(
\Gamma_{B_K} \ln(\mu a)
+ \frac{11}{12} \Gamma_{B_K}
- C^{Latt}_{B_K,2TR}
\right) \right]
\vec{\cal O}^{(1)}_{B_K,2TR}
\ea
where
\ba
\label{projection}
\vec{P}_{\Gamma = -4}\equiv ( 1, 1, 1, 1)
\ea
projects out the desired eigen-operator
$ {\cal O}_{\Gamma = -4} $.
\subsubsection{Lattice One Spin Trace Operators}
Let us consider the following Green's function on the lattice
\ba
\langle \chi(x_1) \chi(y_1) \bar{\chi}(x_2) \bar{\chi}(y_2)
{\cal O }^{Latt}_{SF;SF,1TR} \rangle \ ,
\ea
where $ {\cal O }^{Latt}_{SF;SF,1TR} $ belongs to the one
spin trace formulation defined in section 3.2.
In contrast to the two spin trace operators,
both vector  and  axial channels
require their chiral partner operators
in order to respect the continuum chiral behavior
on the lattice as outlined in section 4.3.
In the one spin trace formalism, we have two
questions to answer.
The first issue is
whether the chiral behavior implemented
by adding a chiral-partner operator is preserved after
renormalization.
The second question is whether
a linear combination of one spin trace operators
can obey the normalization conditions which define the
continuum operator.
This guarantees that
$ R^{Latt}_{ij,SS'} $ in Eq. (\ref{z-2tr})
reproduces the corresponding continuum term
when we make a connection between the lattice and
continuum operators.
The Feynman rules are identical to those of the
two spin trace operators.
Let us restrict our discussion
to Landau gauge operators
such that we only need to
consider 8 diagrams
(Figure 3 (g1), (g2), (g3), (g4)
and Figure 4 (g1), (g2), (g3), (g4))
in addition to the
bilinear diagrams discussed in the previous section
(details of the calculations are
given in appendix D).

The lattice four-fermion operators for $ B_K $
in the one spin trace formalism are given
in Eqs. (\ref{1tr-op}), (\ref{1tr-1})--(\ref{1tr-4})
and their chiral partner operators in Eqs.
(\ref{chi-op}), (\ref{chi-1})--(\ref{chi-4}) respectively.
They belong to the $ I^{++} $ representation
of the Euclidean rotation group
with parity and charge conjugation symmetry \cite{verstegen}.
The mixing of operators is
exactly the same as in the
two spin trace formalism
as described in Eq. (\ref{lattice-one-loop}).
However, in the one spin trace formalism
for $ B_K $, there are eight instead of
four operators which mix with one another
through the anomalous dimension matrix at the one loop level
\ba
\left(
(V  \times P)^{1TR}_{ab;ba}, \
(V  \times P)^{1TR}_{aa;bb}, \
(A  \times P)^{1TR}_{ab;ba}, \
(A  \times P)^{1TR}_{aa;bb} \right)
\nonumber
\\
\mbox{ and } \left(
(A  \times S)^{1TR}_{ab;ba}, \
(A  \times S)^{1TR}_{aa;bb}, \
(V  \times S)^{1TR}_{ab;ba}, \
(V  \times S)^{1TR}_{aa;bb} \right) \ .
\ea

Before we go further to present the explicit results
for the one-loop renormalization, let us consider
the  $ U_A(1) $ symmetry to understand the
mixing of the four-fermion operators
on the lattice better. For Landau gauge operators,
these $ U_A(1) $ rotations
(Eq.(\ref{transf-1}) and
Eq. (\ref{transf-2}, \ref{transf-3}, \ref{transf-4}))
act on each of the bilinears
of the four-fermion operators simultaneously.
{}From these transformations as explained in section 5.2,
we obtain the following
important properties \cite{Japan1,sharpe4}
of four-fermion operator mixing:
\begin{enumerate}
\item[(1)] The distance parity of each bilinear in the
four-fermion operator can not be changed by mixing.
\item[(2)] The radiative corrections for the operators
$
[\bar{\chi} (\overline{\gamma_{S5} \otimes \xi_{F5} }) \chi]
[\bar{\chi} (\overline{\gamma_{S5} \otimes \xi_{F5} }) \chi]
$
are the same as those for
$
[\bar{\chi} (\overline{\gamma_S \otimes \xi_F }) \chi]
[\bar{\chi} (\overline{\gamma_S \otimes \xi_F }) \chi]
$.
\item[(3)] The renormalization coefficients of the
operators
$
[\bar{\chi} (\overline{\gamma_S \otimes \xi_F }) \chi]
[\bar{\chi}
(\overline{\gamma_{S5} \otimes \xi_{F5} }) \chi]
$
can be obtained from those of $
[\bar{\chi} (\overline{\gamma_S \otimes \xi_F }) \chi]
[\bar{\chi} (\overline{\gamma_S \otimes \xi_F }) \chi]
$ by $ U_A(1) $ rotations given
in Eq. (\ref{transf-1}, \ref{transf-2}, \ref{transf-3},
\ref{transf-4}).
\end{enumerate}
Numerically Ishizuka and Shizawa
showed that these results
are true at least at the  one loop level for Landau gauge
operators \cite{Japan1}.
Sharpe and Patel argued
in Ref. \cite{sharpe4}
that these results are valid to all
orders in perturbation theory
for Landau gauge operators.

The property (2) in the above is
useful to determine
the renormalization coefficients of
the one spin trace operators.
This property insures that
the anomalous dimension matrix and
finite mixing terms of $ (V \times P) $
are identical to
the $ (A \times S)
$ by multiplying both bilinears by
$ (\overline{\gamma_5 \otimes \xi_5}) $
and the same is
true for $ (V \times S) $ and $ (A \times P) $.
Therefore,
each four-fermion operator for $ B_K $
has the same radiative correction structure
as its own chiral partner operator.
Hence we can choose an operator basis
(we call it {\em chiral basis})
such that each element satisfies the continuum chiral
behavior and has the same radiative loop corrections
simultaneously. In other words, the operator basis
is composed of
the four-fermion operator plus its chiral partner operator
as follows
\ba
\label{basis}
\vec{\cal{O}}_{B_K,1TR}
\equiv \left( \begin{array}{c}
(V \times P + A  \times S)^{1TR}_{ab;ba} \\
\vspace*{0.01 in} \\
(V \times P + A \times S)^{1TR}_{aa;bb} \\
\vspace*{0.01 in} \\
(A \times P + V  \times S)^{1TR}_{ab;ba} \\
\vspace*{0.01 in} \\
(A \times P + V  \times S)^{1TR}_{aa;bb}
\end{array} \right) \ .
\ea
{}From the standpoint of group theory, this is a projection
of operators to a particular representation of
the $ U_A(1) $ group. The above basis operators
({\em chiral basis operators}) all belong to
the identity representation of a subgroup of $ U_A(1) $
which has the following transformation property:
\ba
\chi_d \longrightarrow \exp[i \frac{\pi}{4}
( \overline{ \gamma_5 \otimes \xi_5} )] \chi_d &,&
\bar{\chi}_d \longrightarrow \bar{\chi}_d
\exp[i \frac{\pi}{4}
( \overline{ \gamma_5 \otimes \xi_5} )]
\\
\chi_s \longrightarrow \exp[-i \frac{\pi}{4}
( \overline{ \gamma_5 \otimes \xi_5} )] \chi_s &,&
\bar{\chi}_s \longrightarrow \bar{\chi}_s
\exp[-i \frac{\pi}{4}
( \overline{ \gamma_5 \otimes \xi_5} )]
\ea
The above transformation is equivalent to choosing
$ \beta = \frac{\pi}{4} $ and $ \eta = - \frac{\pi}{4} $
in Eq. (\ref{transf-3}) and (\ref{transf-4}).

The anomalous dimension matrix for
$ \vec{\cal O}_{B_K,1TR} $ is the same as that for
$ \vec{\cal O}_{B_K,2TR} $
in Eq. (\ref{gamma}) with identical
eigenvalues and corresponding eigen-operators.
The eigen-operator with eigenvalue $ \Gamma = -4 $
is
\ba
{\cal O}_{\Gamma= -4,1TR} & = &
(V \times P + A \times S)^{1TR}_{ab;ba}
+ (V \times P + A \times S)^{1TR}_{aa;bb}
\nonumber
\\
& &
+ (A \times P + V \times S)^{1TR}_{ab;ba}
+ (A \times P + V \times S)^{1TR}_{aa;bb}
\ea
Only one eigen-operator $ {\cal O}_{\Gamma= -4,1TR} $
has the same leading
logarithmic behavior as the
continuum $ \Delta S = 2 $ operator to all orders
in perturbation theory.  In the continuum limit,
the other three operators each contain at least
one of the unphysical quark flavors that we
introduced when defining the lattice theory.
Therefore, these three operators can not contribute
to $ B_K $ and can not be candidates for the
continuum $ \Delta S = 2 $ operator.

Hence by choosing $ {\cal O}_{\Gamma= -4,1TR} $ as the
lattice version of the continuum operator,
we can implement both the
continuum chiral behavior and the
continuum leading logarithmic behavior simultaneously.
It also is guaranteed
that the momentum-dependent finite terms
($ R^{Latt,1TR}_{ij,SS'} $
and $ R_{\Delta S = 2} $) will cancel
when matching
the lattice and continuum
operators at the one loop level.

However,
the scheme-dependent finite terms
$ C^{Latt}_{B_K,1TR} $
are quite different from those in the
two spin trace formalism.
There are two classes of operators which
mix with momentum-independent, and
scale-independent, but scheme-dependent
finite coefficients. The first class,
discussed in this section, is the group of operators
in Eq. (\ref{basis}), which are mixed by the anomalous
dimension matrix at the one loop level.
Other operators, which do not appear in the first class
and  which will turn out to be small later,
make up the second class and are discussed
in the next section.
This finite mixing of the 8 operators
given in Eqs. (\ref{1tr-1})--(\ref{1tr-4})
and Eqs. (\ref{chi-1})--(\ref{chi-4})
needs to be discussed in terms of the
{\em chiral basis operators}
in order to preserve the correct chiral behavior.
Now let us present the finite mixing matrix
$ C^{Latt}_{B_K,1TR} $ in terms of the
basis operator $ \vec{\cal O}_{B_K,1TR} $
given in Eq. (\ref{basis})
\ba
& &
C^{Latt}_{B_K,1TR} =
\left(
\begin{array}{cccc}
 36.908  &  -3.1442  & -4.8246  & -2.394  \\
-0.4612  &  28.860  & -4.5024  &  1.5008  \\
-4.8246  & -2.394   &  37.438  &  -4.7349 \\
-4.5024  &  1.5008  & -0.4612  &  24.618
\end{array}
\right) \ ,
\ea
where the details of calculations are explained
in appendix D.
Now we can connect the lattice operator
$ \vec{\cal O}^{(1)}_{B_K,1TR} $ with the continuum
$ \Delta S = 2 $  operator
$ {\cal O}^{Cont(1)}_{\Delta S = 2} $ as follows
\ba
& &{\cal O}^{(0)}_{\Gamma = -4}  =
\vec{P}_{\Gamma = -4} \cdot \vec{\cal O}^{(0)}_{B_K,1TR}
= {\cal O}^{Cont(0)}_{\Delta S = 2} +
\mbox{ terms of order $ a $}
\nonumber
\\
& &{\cal O}^{Cont}_{\Delta S = 2}
= \vec{P}_{\Gamma = -4} \cdot
\left[ I + \frac{g^2}{(4\pi)^2}
\left(
\Gamma_{B_K} \ln(\mu a)
+ \frac{11}{12} \Gamma_{B_K}
- C^{Latt}_{B_K,1TR}
\right) \right]
\vec{\cal O}_{B_K,1TR}
\nonumber \\
& & \hspace*{30mm} -
\Delta \left(
\vec{P} \cdot \vec{\cal O}_{B_K,1TR}
\right) \ ,
\label{match-1tr}
\ea
where the superscript $ (i) $, $ i = 0,1,2\cdots $
represents the number of loops,
$ \Gamma_{B_K} $ is identical to
Eq. (\ref{gamma}),
$ \vec{P}_{\Gamma = -4} $ is the projection vector
given in Eq. (\ref{projection}) and
$  \Delta ( \vec{P} \cdot \vec{\cal O}_{B_K,1TR} ) $
is the radiative correction which comes from the second class
of operators, discussed in the next section.
\subsubsection{Mixing of Operators with Different Spin-flavor Structure}
Let us discuss the mixing of operators with spin-flavor structure
different from the pseudo-Goldstone Kaon ($ \gamma_5 \otimes \xi_5 $)
in both formalisms.

In the {\em two spin trace formalism},
there are 52 operators which mix with the original
lattice operators in Eq. (\ref{basis}) and
have flavor structure different
from the pseudo-Goldstone Kaon.
Since these operators
need to be contracted
with the external pseudo-Goldstone mode
in the weak matrix elements,
these weak matrix elements are
proportional to a vanishing
flavor trace in the continuum
limit $ a \rightarrow 0 $,
where SU(4) flavor
symmetry is supposed to
be restored.
Thus,
those operators with different flavor structure
($ \xi_F \neq \xi_5 $) are suppressed
by the vanishing flavor trace
(expected to be suppressed by at least one power of $a$)
in addition to the $ g^2/(4\pi)^2 $ suppression
in the {\em two spin trace formalism}.
In fact the only operator that is not
suppressed in this way is the original operator
itself in Eqs. (\ref{op-2tr}--\ref{op-2tr-4}).

In the {\em one spin trace formalism},
a group of operators mixing with the original
operator in Eq. (\ref{basis}) which
has the flavor structure different
from the pseudo-Goldstone Kaon are suppressed
only by the $ g^2/(4\pi)^2 $ factor but not by
the vanishing flavor trace in the contraction with
the external pseudo-Goldstone modes.
This is the reason why we need to take
into account this large group of
remaining operators more carefully in the
{\em one spin trace formalism}.

Let us choose the chiral basis of operators
described in the previous section where the basis operators
belong to identity representation with respect to
the specific $ U_A(1) $ axial rotation transformation.
At the one loop level, our $ ((V+A) \times (P+S))^{1TR} $
are mixed significantly with the
$ (S \times V)^{1TR} + (P \times A)^{1TR} $,
$ (S \times A)^{1TR} + (P \times V)^{1TR} $,
$ (V \times T)^{1TR} + (A \times T)^{1TR} $ operators
which have non-negligible matrix elements
with the pseudo-Goldstone Kaons:
\ba
& & \vec{P} \cdot \vec{\cal O}_{B_K,1TR} =
\sum'\ \left( (V+A) \times P
+ (V+A) \times S \right)^{(1)1TR} \ ,
\nonumber \\
& & \Delta \left(
\vec{P} \cdot \vec{\cal O}_{B_K,1TR}
\right) \equiv
\nonumber \\
& & \hspace*{10mm}
\frac{g^2}{(4\pi)^2} \left[
-1.214 \left(S \times V + P \times A \right)^{(0)1TR}_{ab;ba}
-2.805 \left(S \times V + P \times A \right)^{(0)1TR}_{aa;bb}
\right.
\nonumber \\
& & \hspace*{25mm}
-4.13 \left(S \times A + P \times V \right)^{(0)1TR}_{ab;ba}
-2.539 \left(S \times A + P \times V \right)^{(0)1TR}_{aa;bb}
\nonumber \\
& &  \hspace*{25mm}
-0.402 \left(V \times T + A \times T \right)^{(0)1TR}_{ab;ba}
-2.228 \left(V \times T + A \times T \right)^{(0)1TR}_{aa;bb}
\nonumber \\
& & \hspace*{25mm} \left.
+ \cdots \right] \  ,
\label{op-mix}
\ea
where $ \sum' $ means summation over two kinds of color
contraction. Here we have written only those finite terms
corresponding to the wrong flavor operators,
terms that correspond to
$ \Delta ( \vec{P} \cdot \vec{\cal O}_{B_K,1TR} ) $
in Eq. (\ref{match-1tr}).

The other operators ($\cdots$ in Eq. (\ref{op-mix}))
mixing at the one loop level,
which are not written explicitly
in Eq. (\ref{op-mix}),
are negligible
({\em i.e.} expected to be of order $ a $)
due to the vanishing flavor trace
in the matrix element with
the pseudo-Goldstone Kaon.
They are
\ba
& &
(T \times V)^{1TR} + (T \times A)^{1TR} \ ,
\label{op-tensor}
\\
& &
\sum_{\mu \neq \nu} \left[
\bar{\chi}_{\rm s}
(\overline{ I \times \xi_{\mu} }) \chi_{\rm d} \
\bar{\chi}_{\rm s'}
(\overline{ I \times \xi_{\nu} }) \chi_{\rm d'}
\  + \
\bar{\chi}_{\rm s}
(\overline{ \gamma_5 \times \xi_{\mu5} }) \chi_{\rm d} \
\bar{\chi}_{\rm s'}
(\overline{ \gamma_5 \times \xi_{\nu} }) \chi_{\rm d'}
\right] ,
\label{op-mix-1}
\\
& &
\sum_{\mu \neq \nu} \left[
\bar{\chi}_{\rm s}
(\overline{ \gamma_5 \times \xi_{\mu} }) \chi_{\rm d} \
\bar{\chi}_{\rm s'}
(\overline{ \gamma_5 \times \xi_{\nu} }) \chi_{\rm d'}
\  + \
\bar{\chi}_{\rm s}
(\overline{ I \times \xi_{\mu5} }) \chi_{\rm d} \
\bar{\chi}_{\rm s'}
(\overline{ I \times \xi_{\nu} }) \chi_{\rm d'}
\right] ,
\label{op-mix-2}
\\
& & \hspace*{-3mm}
\sum_{\mu, \sigma \neq \nu \neq \rho} \left[
\bar{\chi}_{\rm s}
(\overline{ \gamma_{\mu} \times \xi_{\sigma \nu} }) \chi_{\rm d} \
\bar{\chi}_{\rm s'}
(\overline{ \gamma_{\mu} \times \xi_{\sigma \rho} }) \chi_{\rm d'}
\  +  \
\bar{\chi}_{\rm s}
(\overline{ \gamma_{\mu} \times \xi_{\sigma \nu} }) \chi_{\rm d} \
\bar{\chi}_{\rm s'}
(\overline{ \gamma_{\mu} \times \xi_{\sigma \rho} }) \chi_{\rm d'}
\right] ,
\label{op-mix-3}
\\
& &
\sum_{\mu \neq \nu} \left[
\bar{\chi}_{\rm s}
(\overline{ \gamma_{\mu \nu} \times \xi_{\mu} }) \chi_{\rm d} \
\bar{\chi}_{\rm s'}
(\overline{ \gamma_{\mu \nu} \times \xi_{\nu} }) \chi_{\rm d'}
\right]
\nonumber \\
& & \hspace*{30mm}  +
\sum_{\mu \neq \sigma \neq \nu \neq \rho} \left[
\bar{\chi}_{\rm s}
(\overline{ \gamma_{\mu \nu} \times \xi_{\sigma 5} }) \chi_{\rm d} \
\bar{\chi}_{\rm s'}
(\overline{ \gamma_{\mu \nu} \times \xi_{\rho 5} }) \chi_{\rm d'}
\right],
\label{op-mix-4}
\\
& & \hspace*{-3mm}
\sum_{\mu \neq \nu \neq \sigma \neq \rho} \left[
\bar{\chi}_{\rm s}
(\overline{ \gamma_{\mu \nu} \times \xi_{\sigma} }) \chi_{\rm d} \
\bar{\chi}_{\rm s'}
(\overline{ \gamma_{\mu \nu} \times \xi_{\rho} }) \chi_{\rm d'}
\right]
\nonumber \\
& & \hspace*{30mm}  +
\sum_{\mu \neq \nu} \left[
\bar{\chi}_{\rm s}
(\overline{ \gamma_{\mu \nu} \times \xi_{\mu 5} }) \chi_{\rm d} \
\bar{\chi}_{\rm s'}
(\overline{ \gamma_{\mu \nu} \times \xi_{\nu 5} }) \chi_{\rm d'}
\right],
\label{op-mix-5}
\ea

In principle, the presence of the operators
in Eq. (\ref{op-mix}) causes an extra difficulty
for the one spin trace formalism, since there
are now additional operators whose matrix elements
must be computed.

However,
in order to evaluate the new matrix elements
we can analytically express these
one spin trace operators in Eq. (\ref{op-mix})
in terms of the small conventional set of
two spin trace operators by Fierz transformation.
Since only those operators
with the same flavor structure ($ \xi_5 $)
as the pseudo-Goldstone mode are significant and
the other operators are suppressed by the
vanishing flavor trace
in the {\em two spin trace formalism},
it is possible to calculate approximately
each one spin trace operator
in Eq. (\ref{op-mix})
from the numerical results of the
two spin trace operator measurements on the lattice.
Thus the class of required matrix elements in the
one spin trace formalism need not be expanded
further \cite{los}.

Therefore, only those one spin trace operators whose
Fierz transform
has a non-vanishing component of either
$ (V \times P )^{2TR} $ or $ (A \times P)^{2TR} $
are important.
Now let us figure out which operators
in Eq. (\ref{op-mix}) and
Eqs. (\ref{op-mix-1}--\ref{op-mix-5})
can contribute to $ B_K $.
The Fierz-transforms of the operators in
Eq. (\ref{op-mix}) and Eq. (\ref{op-mix-1}) are
\ba
& & (S \times V)^{1TR} + (P \times A)^{1TR}
\nonumber \\
& & =  \frac{1}{2} \left(
(S -T +P) \times (V + A) \right)^{2TR}
- \left(
(V - A ) \times ( S - P ) \right)^{2TR} \  ,
\\
& & ( S \times A)^{1TR} + (P \times V)^{1TR}
\nonumber \\
& & = \frac{1}{2} \left(
(S -T +P) \times (V + A) \right)^{2TR}
+ \left(
(V - A ) \times ( S - P ) \right)^{2TR} \ ,
\\
& & \left( (V + A) \times T \right)^{1TR}
\nonumber \\
& & = - 3
\left( ( V + A ) \times (3S + T + 3P) \right)^{2TR} \ ,
\\
& & \left( T \times (V + A) \right)^{1TR} \ ,
\nonumber \\
& & =
- \left( ( 3S + T + 3P ) \times (V + A) \right)^{2TR}
\ .
\ea
If we neglect those two spin trace operators with
the wrong flavor structure ($ \xi_F \neq \xi_5 $),
the above equations can be simplified as follows:
\ba
& & (S \times V)^{1TR} + (P \times A)^{1TR}
\longrightarrow
\left( (V - A ) \times P \right)^{2TR} \  ,
\\
& & ( S \times A)^{1TR} + (P \times V)^{1TR}
\longrightarrow
- \left( (V - A ) \times P \right)^{2TR} \ ,
\\
& & \left( (V + A) \times T \right)^{1TR}
\longrightarrow
-3 \left( ( V + A ) \times P \right)^{2TR}\  ,
\\
& & \left( T \times (V + A) \right)^{1TR}
\longrightarrow 0 \ .
\ea
The Fierz transforms of the operators in Eqs.
(\ref{op-mix-1}--\ref{op-mix-5}) contain
neither $ (V \times P)^{2TR} $ nor
$ (A \times P)^{2TR} $ since those operators
have non-diagonal flavor structure ({\em i.e.}
the flavor matrix of one bilinear is different
from that of the other bilinear) and they can not be
Fierz-transformed into a desired diagonal flavor structure.
Therefore, we conclude that the operators
in Eq. (\ref{op-mix}) do contain
$ (V \times P )^{2TR} $ and  $ (A \times P)^{2TR} $
but the operators in Eqs. (\ref{op-mix-1}--\ref{op-mix-5})
include neither
$ (V \times P )^{2TR} $ nor $ (A \times P)^{2TR} $.

Now we can express Eq. (\ref{op-mix})
in terms of the non-trivial
two spin trace operators as follows:
\ba
\Delta \left(
\vec{P} \cdot \vec{\cal O}_{B_K,1TR} \right)
& = &
\frac{g^2}{(4\pi)^2} \left[
4.122 (V \times P)^{2TR}_{aa;bb}
+6.418 (V \times P)^{2TR}_{ab;ba} \right.
\nonumber \\
& & \hspace*{20mm} \left.
-1.710 (A \times P)^{2TR}_{aa;bb}
+6.950 (A \times P)^{2TR}_{ab;ba}
\right] \ .
\label{op-mix-fine}
\ea
The equation (\ref{op-mix-fine}) can tell us
how much of the radiative corrections
at the one loop level are neglected
in the leading term of
the matching formula Eq. (\ref{match-1tr}).
\subsection{Tadpole Improvement}
Parisi, Lepage and Mackenzie showed that the tadpole diagrams
are the main source of the large difference between the
bare lattice coupling $ g_{0}(a) $ and the renormalized
coupling $ g_{\overline{MS}}( \mu = \frac{1}{a} ) $
\cite{parisi,lepage,sharpe3,Japan1}.
They suggested a mean field method for removing the
dominant effect of tadpole diagrams.
Noticing that the vacuum expectation value of the
link matrix is smaller than 1, they proposed that the
appropriate connection with the continuum gauge field is
\ba
U_{\mu}(x) \rightarrow u_{0}(1 + iagA_{\mu}) \ ,
\ea
where $ u_0 $ represents the mean value of the link matrix.
One possible choice of a gauge-invariant $ u_0 $
is
\ba
u_0 \equiv \left[ \mbox{Re} \langle \frac{1}{3}
\mbox{Tr} U_{\Box} \rangle \right]^{\frac{1}{4}}
\ .
\ea
It is the rescaled link $ \frac{U_{\mu}}{u_0} $
that should be expanded around unity. The staggered
fermion action can be rewritten in terms of
the rescaled fields
$ \psi \equiv \sqrt{u_0} \chi $
\ba
S_{\mbox{\small fermion}}
& = & - \sum_{x,\mu} \frac{1}{2} \eta_{\mu}(x)
\left[
\bar{\psi}(x)
\frac{ U_{\mu}(x) }{u_0}
\psi(x+a_{\mu}) -
\bar{\psi}(x+a_{\mu})
\frac{ U_{\mu}^{\dagger}(x) }{u_0}
\psi(x)
\right]
\nn \\
& & \hspace*{20mm}
- \frac{M}{u_0} a \sum_x
\bar{\psi}(x) \psi(x) \ ,
\ea
where we take zero quark mass ($ M = 0 $) for the renormalization
and the gluon action is
\ba
S_{\mbox{\small gluon}} & = & -\sum_{\Box} \beta_{MF}
\frac{1}{N_c}
\frac{ \mbox{ReTr}U_{\Box} }{ u^{4}_0 } \ ,
\ea
where
\ba
\beta_{MF} = \frac{2N_c}{g^{2}_{MF}},
\  \mbox{ with }
g^{2}_{MF} = \frac{g^{2}_0(a)}
{ \mbox{Re} \langle \frac{1}{3}
\mbox{Tr} U_{\Box} \rangle } \ .
\ea
It has been found out that $ \psi $
matches the continuum quark field
better than $ \chi $.
The perturbative expansion for Tr$U_{\Box}$ is given
in Ref. \cite{karsch}
\ba
\mbox{Re} \langle \frac{1}{3}
\mbox{Tr} U_{\Box} \rangle = 1 - \frac{1}{3} g^2_{0}(a) +
O(g^{4}_0(a)) \ .
\ea
$ g^{2}_{MF} $ is related
to $ g^{2}_{0} $ and
$ g^{2}_{\overline{MS}} $ perturbatively
as follows
\ba
g^{2}_{MF} & = &
g^{2}_{0}(a)
\left[1 + \frac{1}{3} g^{2}_{0}(a) +
O(g^{4}_0(a)) \right]
\\
g^{2}_{ \overline{MS}}(\mu) & = &
g^{2}_{MF}
\left[ 1-\beta_{0} g^{2}_{MF}
\{
2 \ln(a\mu) + 2 \ln(\frac{ \Lambda_{L} }
{ \Lambda_{\overline{MS}} }) \}
\right.
\nonumber
\\
& &
\left.
-\frac{1}{3}g^{2}_{MF} + O(g^{4}_{MF})
\right] \ .
\ea
where the details of the derivation are given in Ref.
\cite{wlee}.

Now let us think about the tadpole improvement
for the bilinear and four-fermion operators.
For these operators,
the tadpole contribution can be
removed by rescaling the fermion fields and
gauge links as follows
\ba
\chi \rightarrow \psi =\sqrt{u_0} \chi
\\
\bar{\chi} \rightarrow \bar{\psi} = \sqrt{u_0}
\bar{\chi}
\\
U_{\mu} \rightarrow \frac{ U_{\mu} }{ u_0} \ .
\ea
The tadpole improvement for the bilinear
Landau gauge operators is therefore
\ba
{\cal O}^{MF}_{SF} = u_0
{\cal O}_{SF} \ ,
\ea
whereas
for the four-fermion Landau gauge
operators we have
\ba
{\cal O}^{MF}_{SF;S'F'} =
(u_0)^{2}{\cal O}_{SF;S'F'} \ .
\ea
The only effect of tadpole improvement for the Landau
gauge operators is that they are rescaled by a
gauge invariant factor.
As a consequence, the scheme-dependent finite terms
change to
\ba
C^{Latt,MF}_{B_K,2TR} & = &
\left(
\begin{array}{cccc}
11.207	& -2.9136 & -5.25285 & -2.2512  \\
0	& 2.387  & -4.5024  & 1.5008   \\
-5.25285& -2.2512 & 11.657  & -4.5043  \\
-4.5024 & 1.5008  &  0	    & -1.855
\end{array}
\right)
\\
C^{Latt,MF}_{B_K,1TR} & = &
\left(
\begin{array}{cccc}
 10.589  &  -3.1442  & -4.8246  & -2.394   \\
-0.4612  &  2.541   & -4.5024  &  1.5008  \\
-4.8246  & -2.394   &  11.119  &  -4.7349  \\
-4.5024  &  1.5008  & -0.4612  &  -1.701
\end{array}
\right) \ .
\ea
We would like to point out that
for tadpole improvement,
one has to use the renormalized
coupling $ g^{2}_{\overline{MS}}(\mu = \frac{1}{a}) $
or $ g^{2}_{MF} $
rather than the lattice bare coupling
$ g^{2}_0(a) $ \cite{lepage,wlee}.
\section{Summary and Conclusions}
We have presented two different methods to transcribe
the continuum  $ \Delta S = 2 $ operator onto the
lattice: one is the {\em two spin
trace formalism} (conventional method) and
the other is the {\em one spin trace
formalism}.

The chiral behavior of the continuum $ \Delta S = 2 $
operator needs to be respected by the lattice operators.
Once the lattice operators are chosen, it is also important
to check whether
the lattice operators preserve the leading logarithmic
behavior of the continuum $ \Delta S = 2 $ operator.
Since both the lattice and the continuum operators
have the same leading logarithmic behavior,
we are guaranteed that
those terms which depend on the external momenta cancel
when we match the lattice measurement to its continuum
correspondence.

In the {\em two spin trace formalism},
it is shown that
vector and axial operators satisfy
the continuum chiral behavior
separately.
In the {\em one spin trace formalism},
each operator requires an additional chiral partner operator
in order to obey the correct continuum chiral behavior.
Hence 8 four-fermion operators are requisite for $ B_K $
in the one spin trace formalism while only 4 four-fermion
operators suffice for $ B_K $ in the
two spin trace formalism.
Furthermore, in the {\em one spin trace formalism}
the basis operators are chosen such that
they belong to the identity representation
of the $ U_A(1) $ symmetry group.
In this representation, we can find an eigen-operator
which has the same chiral behavior and the same
leading logarithmic behavior as the continuum
$ \Delta S = 2 $ operator.
Therefore, in both formalisms
we find an eigen-operator which has the same
leading logarithmic behavior as the continuum
$ \Delta S = 2 $ operator.

However, the scheme-dependent finite terms differ
in the two formalisms, because the 8 operators
mix with one another in the one spin trace formalism
while the 4 operators in the two spin trace formalism.
In addition, tadpole improvement through mean field theory
is important to improve the matching between the
lattice calculation and the continuum observable ($B_K$).

The lattice operators for $B_K$
in the one spin trace formalism can not
be Fierz transformed into the corresponding operators
for $B_K$ in the two spin trace formalism.
However, in the continuum limit of $ a \rightarrow 0 $
the contributions of the different flavor structures
(for example $ ((V+A)\times S)^{2TR} $
or $ ((V+A)\times T)^{2TR} $)
vanish away. This guarantees that in the continuum
limit $a \rightarrow 0$, the matrix elements of the operators
with the external pseudo-Goldstone bosons
agree in both formalisms.

We intend to
apply both formalisms in our numerical simulations
of lattice QCD. We will match the lattice observables
with the continuum correspondence through the
one loop relation given in this paper and will
compare the results of both formalisms.
We believe that this comparison will tell us
how close our numerical simulation is to
the continuum physics.
\section*{Acknowledgements}
We would like to express our gratitude to
Prof. Norman H. Christ and Prof. Robert D. Mawhinney
for their helpful discussions. Helpful conversations
with Prof. S. Sharpe, Prof. G. Kilcup, Dr. R. Gupta
and Prof. A. Ukawa
during the Santa Fe meeting (July, 1994) held
by Los Alamos National Laboratory
as well as conversations with Dr. Apoorva Patel
are acknowledged with gratitude.
We appreciate Dr. Ishizuka's kind help
for providing the finite integral tables to us.
Careful proof-reading of this article by Adrian
Kaehler is also acknowledged with gratitude.
\newpage
\appendix
\begin{center}
{\Large \bf Appendix}
\end{center}
\section{Lattice Ward Identity}
Here, we derive a useful Ward identity
on the lattice which will be used to
explain the chiral behavior of four-fermion
operators.
This Ward identity appeared originally
in Ref. \cite{tool-kit}
but with minor typing errors in the formula.

The staggered fermion propagator satisfies the following
relation
\ba
[D+m] G(n_{1},n_{2}) &=& \delta_{n_{1},n_{2}}
\\
< n_{1} \mid D \mid n_{2} >
&\equiv& \sum_{\mu} \frac{1}{2} \eta_{ \mu }( n_{1} )
\tilde{U}(n_{1},n_{2})
\{ \delta_{n_{2},n_{1} + {\mu}} -
\delta_{n_{2},n_{1} - {\mu}}
\}
\nn \\
&\equiv& \half U(n_{1},n_{2})
\\
G(n_{1},n_{2}) &\equiv& < 0 \mid
T[ \chi(n_{1}) \bar{\chi}(n_{2}) ]
\mid 0 >
\ea
Using a $ \frac{1}{m} $ expansion, we can obtain
another representation of $ G(n_{1},n_{2}) $
\ba
G(n_{1},n_{2})
& = & < n_{1} \mid (D+m)^{-1} \mid n_{2} >
\nn \\
& = & \frac{1}{m} \sum_{n}
< n_{1} \mid (-\frac{D}{m} )^{n} \mid n_{2} >
\nn \\
\label{m-exp}
& = & \sum_{\Gamma(0) = n_{1},
\Gamma(\mid \Gamma \mid) = n_{2}}
(-1)^{n_{1}-n_{2}} (\frac{1}{m})^{\mid \Gamma \mid + 1}
(\frac{1}{2})^{\mid \Gamma \mid}
U_{\Gamma}(n_{1},n_{2})
\ea
Since $ U_{\Gamma} $ is unitary,
$ U_{\Gamma_{1}\Gamma_{2}} =
U_{\Gamma_{1}}U_{\Gamma_{2}} $ is also unitary.

Now let us consider the following Green's function
\ba
S(n_{1},n_{2}) \equiv \sum_n G_1(n_1;n)
\epsilon(n) G_2(n;n_2)
\ea
Using Eq. (\ref{m-exp}),
$ S(n_{1},n_{2}) $ can be simplified
as follows
\ba
S(n_{1},n_{2}) & = &
\sum_{n} \sum_{\Gamma_{1}}\sum_{\Gamma_{2}}
(-1)^{n_{1}-n} (-1)^{n-n_{2}}
(\frac{1}{m_1})^{\mid \Gamma_{1} \mid + 1}
(\frac{1}{m_2})^{\mid \Gamma_{2} \mid + 1}
\nn \\
& & \hspace{0.5 in}
(\frac{1}{2})^{\mid \Gamma_{1} \mid +
\mid \Gamma_{2} \mid}
U_{\Gamma_{1}}(n_{1},n)
U_{\Gamma_{2}}(n,n_{2}) \epsilon(n)
\nn \\
& = &
\sum_{\Gamma,\Gamma(0)=n_1,
\Gamma(\mid \Gamma \mid) = n_2}
(-1)^{n_{1}-n_{2}}
(\frac{1}{2})^{\mid \Gamma \mid}
U_{\Gamma}(n_{1},n_{2})
\nn \\
& & \hspace{0.5 in}
\sum_{i=0}^{\mid \Gamma \mid}
(-1)^{\Gamma(0)} (-1)^{i}
(\frac{1}{m_1})^{ i + 1}
(\frac{1}{m_2})^{\mid \Gamma \mid -i + 1}
\nn \\
& = & \sum_{\Gamma} (-1)^{n_{1}-n_{2}}
(\frac{1}{2})^{\mid \Gamma \mid}
U_{\Gamma}(n_{1},n_{2})
(-1)^{\Gamma(0)}
\nn \\
& & \hspace{0.5 in}
\frac{1}{m_1+m_2}
\{
(\frac{1}{m_2})^{\mid \Gamma \mid + 1}
+ (-1)^{\mid \Gamma \mid }
(\frac{1}{m_1})^{\mid \Gamma \mid + 1}
\}
\nn \\
\label{ward-1}
& = &
\frac{1}{m_1+m_2}
\{
\epsilon(n_{1}) G_{2}(n_{1},n_{2})
+G_{1}(n_{1},n_{2}) \epsilon(n_{2})
\}
\ea
{}From Eq. (\ref{ward-1}), we can derive the following
Ward identity
\ba
\label{ward-2}
(m_1+m_2) \sum_n G_1(n_1;n) \epsilon(n) G_2(n;n_2)
=
\epsilon(n_{1}) G_{2}(n_{1},n_{2})
+G_{1}(n_{1},n_{2}) \epsilon(n_{2})
\ea
Equation (\ref{ward-2})
will be used later for the analysis of
the chiral behavior on the lattice.
\section{Ward Identities for the Two Spin Trace Formalism}
We are interested in the following correlation function
\be
M_K^{2TR}(t,t') = \langle K(t) {\cal O}_{2TR} K'(t') \rangle \;,
\ee
The relevant two spin trace operator is given by
\be
	{\cal O}_{2TR}(y) = \frac{1}{N_f^2}
	[\chibs  (y_A) \o{\gs}{\xf}{AB} \chid  (y_B)]
	[\chibsp (y_C) \o{\gs}{\xf}{CD} \chidp (y_D)] \;.
\ee
Here hypercubes are labeled by capital roman letters
and there is always
(unless specified otherwise)
an implicit sum over repeated indices.
We define $ y_A \equiv 2y + A $ with $ A \in \{0,1 \}^4 $.
The Fourier transform $\ckk$ of $M_K^{2TR}$,
\be
	\ckk^{2TR}(E,E') = \sum_{t,t'}
	\exp(iEt) M_K^{2TR}(t,t') \exp(-iE't') \;,
\ee
%
%
is parameterized as in the Ref. \cite{sharpe1}.
Let us evaluate
\ba
	\ckk^{2TR}(0,0)
    &=& \sum_{t,t'} M_K^{2TR}(t,t') \\
    &=& \bxv{
	\sum_{y,y'} \;
	[\chibd (y) \epsilon(y) \chis (y)]
	[\chibs (0_C) \o{\gs }{\xf }{CD} \chid (0_D)] \nn \\
    & & \hspace{1.4cm}
	[\chibsp (0_E)  \o{\gs }{\xf }{EF} \chidp (0_F) ]
	[\chibdp (y') \epsilon(y') \chi_{s'} (y')]
	} \nn \\
    &=& \sum_{y,y'}
	G_{\rm d }(0_D;y) \epsilon(y) G_{\rm s }(y;0_C)
	\o{\gs}{\xf}{CD} \nn \\
    & &	\hspace{1.2cm}
	G_{\rm d'}(0_F;y') \epsilon(y') G_{\rm s'}(y';0_E)
	\o{\gs}{\xf}{EF} \nn \\
    &=& \frac{1}{m_{s }+m_{d }}\biggl\{
	G_{\rm d }(0_D;0_C)\epsilon(C)
	+ \epsilon(D) G_{\rm s }(0_D;0_C)
	\biggr\} \o{\gs}{\xf}{CD} \nn \\
    & & \frac{1}{m_{s'}+m_{d'}}\biggl\{
	G_{\rm d'}(0_F;0_E)\epsilon(E)
	+\epsilon(F) G_{\rm s'}(0_F;0_E)
	\biggr\} \o{\gs}{\xf}{EF} \nn
\ea
Already at this point it is obvious that for $m_s=m_d$ and for
$m_{s'}=m_{d'}$
the above expression vanishes for odd distance operators,
i.e. $|C-D|=|E-F|=\mbox{odd}$.
But let us continue
\ba
    &=& \frac{1}{m_{s }+m_{d }}\biggl\{
	G_{\rm d }(0_D;0_C) \o{\g_5\gs}{\x_5\xf}{CD} +
	G_{\rm s }(0_D;0_C) \o{\gs\g_5}{\xf\x_5}{CD}
	\biggr\} \nn \\
    & & \frac{1}{m_{s'}+m_{d'}}\biggl\{
	G_{\rm d'}(0_F;0_E) \o{\g_5\gs}{\x_5\xf}{EF} +
	G_{\rm s'}(0_F;0_E) \o{\gs\g_5}{\xf\x_5}{EF}
	\biggr\} \nn \\
    &=& \bxv{
	\frac{1}{m_{s }+m_{d }} \biggl\{
	\chibd (0_C) \o{\g_5\gs}{\x_5\xf}{CD} \chid (0_D) +
	\chibs (0_C) \o{\gs\g_5}{\xf\x_5}{CD} \chis (0_D)
	\biggr\} \nn \\
    & &	\frac{1}{m_{s'}+m_{d'}}	\biggl\{
	\chibdp (0_E) \o{\g_5\gs}{\x_5\xf}{EF} \chidp (0_F) +
	\chibsp (0_E) \o{\gs\g_5}{\xf\x_5}{EF} \chisp (0_F)
	\biggr\}
	} \nn \\
    &=& \frac{1}{m_{s }+m_{d }} \; \frac{1}{m_{s'}+m_{d'}}
	\bxv{ \op_s + \op_d }
    \equiv C_{\op\op} \nn
\ea
We see that the expression vanishes for $m_s=m_d$ and for
$m_{s'}=m_{d'}$ if
\ba
	\g_5\gs &=& \mp \gs\g_5  \mbox{ and} \\
	\x_5\xf &=& \pm \xf\x_5  \nn \;.
\ea
This holds seperately for the $ (V \times P)$
and $ (A \times P) $ channel,
\ba
	\o{\gs}{\xf}{} &=& \o{\gmu}{\x_5}{}   \;,\\
	\o{\gs}{\xf}{} &=& \o{\gmu\g_5}{\x_5}{} \;,
\ea
where the flavor is fixed to $ \xi_5 $ as explained
in section 3.1.
Notice that the vector channel is a distance-three
operator
while the axial channel is a distance-one operator.
Clearly, the correct chiral limit is obeyed
by each channel independently\cite{sharpe1}.
\section{Ward Identities for the One Spin Trace Formalism}
Now we are interested in the following correlation function
\be
M_K^{1TR}(t,t') = \langle K(t) {\cal O}_{1TR} K'(t') \rangle \ .
\ee
The relevant one spin trace operator is given by
\be
	{\cal O}_{1TR}(y) = \frac{1}{N_f}
	[\chibs  (y_A) \o{\gs}{\xf}{AB} \chidp  (y_B)]
	[\chibsp (y_C) \o{\gs}{\xf}{CD} \chid (y_D)] \;.
\ee
Now we repeat the above calculation for the one spin trace case.
We have
\be
	\ckk^{1TR}(E,E') = \sum_{t,t'}
	\exp(iEt) M_K^{1TR}(t,t') \exp(-iE't') \ ,
\ee
and
\ba
\ckk^{1TR}(0,0)
    &=& \sum_{t,t'} M_K^{1TR}(t,t') \\
    &=& \bxv{
	\sum_{y,y'} \;
	[\chibd  (y) \epsilon(y) \chis (y)]
	[\chibsp (0_C) \o{\gs }{\xf }{CD} \chid (0_D)] \nn \\
    & & \hspace{1.4cm}
	[\chibs  (0_E)  \o{\gs }{\xf }{EF} \chidp (0_F) ]
	[\chibdp (y') \epsilon(y') \chisp (y')]
	} \nn \\
    &=& -\sum_{y,y'}
	G_{\rm  d }(0_D;y) \epsilon(y) G_{\rm s }(y;0_E)
	\o{\gs}{\xf}{CD} \nn \\
    & &	\hspace{1.2cm}
	G_{\rm  d'}(0_F;y') \epsilon(y') G_{\rm  s'}(y';0_C)
	\o{\gs}{\xf}{EF} \nn \\
    &=& -\frac{1}{m_{s }+m_{d }}\biggl\{
	G_{\rm  d }(0_D;0_E)\epsilon(E)
	+ \epsilon(D) G_{\rm  s }(0_D;0_E)
	\biggr\} \o{\gs}{\xf}{CD} \nn \\
    & & \cdot \frac{1}{m_{s'}+m_{d'}}\biggl\{
	G_{\rm  d'}(0_F;0_C)\epsilon(C)
	+ \epsilon(F) G_{\rm  s'}(0_F;0_C)
	\biggr\} \o{\gs}{\xf}{EF} \nn
\ea
Contrary to the two spin trace case, at this point it is not evident
that the above expression vanishes
for $m_s=m_d$ or for $m_{s'}=m_{d'}$,
because the distances between
$D$ and $E$ and between $F$ and $C$ are not fixed.
\ba
\ckk^{1TR}(0,0)
    &=&-\frac{1}{m_{s }+m_{d }}
	\frac{1}{m_{s'}+m_{d'}} \nn \\
    & &	\biggl\{
    	G_{\rm d }(0_D;0_E) G_{\rm d'}(0_F;0_C)
	\o{\g_5\gs}{\x_5\xf}{EF} \o{\g_5\gs}{\x_5\xf}{CD} + \nn \\
    & &	\hspace{0.3cm}
	G_{\rm d }(0_D;0_E) G_{\rm s'}(0_F;0_C)
	\o{\g_5\gs\g_5}{\x_5\xf\x_5}{EF} \o{\gs}{\xf}{CD} + \nn \\
    & &	\hspace{0.3cm}
	G_{\rm s }(0_D;0_E) G_{\rm d'}(0_F;0_C)
	\o{\gs}{\xf}{EF} \o{\g_5\gs\g_5}{\x_5\xf\x_5}{CD} + \nn \\
    & &	\hspace{0.3cm}
	G_{\rm s }(0_D;0_E) G_{\rm s'}(0_F;0_C)
	\o{\gs\g_5}{\xf\x_5}{EF} \o{\gs\g_5}{\xf\x_5}{CD}
	\biggr\} \nn \\
    &=& \frac{1}{m_{s }+m_{d }}
	\frac{1}{m_{s'}+m_{d'}} \nn \\
    & &	\bxv{
	[\chibd(0_E)  \o{\g_5\gs}{\x_5\xf}{EF} \chidp(0_F)]
	[\chibdp(0_C) \o{\g_5\gs}{\x_5\xf}{CD} \chid(0_D)] + \nn \\
    & &	\hspace{0.3cm}
	[\chibd(0_E)  \o{\g_5\gs\g_5}{\x_5\xf\x_5}{EF} \chisp(0_F)]
	[\chibsp(0_C) \o{\gs}{\xf}{CD} \chid(0_D)] + \nn \\
    & &	\hspace{0.3cm}
	[\chibs(0_E)  \o{\gs}{\xf}{EF} \chidp(0_F)]
	[\chibdp(0_C) \o{\g_5\gs\g_5}{\x_5\xf\x_5}{CD} \chis(0_D)] + \nn \\
    & &	\hspace{0.3cm}
	[\chibs(0_E)  \o{\gs\g_5}{\xf\x_5}{EF} \chisp(0_F)]
	[\chibsp(0_C) \o{\gs\g_5}{\xf\x_5}{CD} \chis(0_D)]
	} \nn
\ea
We see that the expression would vanish for $m_s=m_d$
and separately for $m_{s'}=m_{d'}$ if
\ba
\gs &=& \sig_1\g_5\gs = \sig_2\gs\g_5  \\
\xf &=& \sig_3\x_5\xf = \sig_4\xf\x_5  \;,
\ea
where $\sig$ is one of the following eight combinations
\ba
\sig = &&(-+++),(+-++),(++-+),(+++-), \nn \\
       &&(+---),(-+--),(--+-),(---+)      \;.
	\label{eq-sig}
\ea
However, this condition can not be satisfied
for $\gs,\xf \in \{I,\xi_{\mu},\sigmunu,\xi_5,\xi_{\mu5}\}$.
Instead, one requires an additional channel
(which we call {\em chiral partner})
with different spin and flavor $ S' $, $ F'$ in order
to achieve the correct chiral limit. The contribution is
\ba
    C'^{1TR}_{KK'}(0,0)
    &=& - \frac{1}{m_{s }+m_{d }} \;
	\frac{1}{m_{s'}+m_{d'}}  \\
    & &	\biggl\{
    	G_{\rm d }(0_D;0_E) G_{\rm d'}(0_F;0_C)
	\o{\g_5\gsp}{\x_5\xfp}{EF} \o{\g_5\gsp}{\x_5\xfp}{CD} + \nn \\
    & &	\hspace{0.3cm}
	G_{\rm d }(0_D;0_E) G_{\rm s'}(0_F;0_C)
	\o{\g_5\gsp\g_5}{\x_5\xfp\x_5}{EF} \o{\gsp}{\xfp}{CD} \, +
	\biggr. \nn \\
    & &	\biggl.
	\hspace{0.3cm}
	G_{\rm s }(0_D;0_E) G_{\rm d'}(0_F;0_C)
	\o{\gsp}{\xfp}{EF} \o{\g_5\gsp\g_5}{\x_5\xfp\x_5}{CD} + \nn \\
    & &	\hspace{0.3cm}
	G_{\rm s}(0_D;0_E) G_{\rm s'}(0_F;0_C)
	\o{\gsp\g_5}{\xfp\x_5}{EF} \o{\gsp\g_5}{\xfp\x_5}{CD}
	\biggr\} \nn \;.
\ea
We want the sum $\ckk^{1TR} + C'^{1TR}_{KK'}$ to vanish
for $m_s=m_d$ and for $m_{s'}=m_{d'}$.
For $m_s=m_d$, the first term in the unprimed channel
will have to cancel the third term in the primed channel,
and vice versa. Similarly, the second and fourth
terms will cancel for $m_{s'}=m_{d'}$.
Demanding that
$\ckk^{1TR} + C'^{1TR}_{KK'}$  in the degenerate mass limit,
one thus obtains the following conditions
\ba
	\gs  &=& \sig_1 \g_5\gsp    \\
	\gsp &=& \sig_2 \gs \g_5    \\
	\xf  &=& \sig_3 \x_5\xfp    \\
	\xfp &=& \sig_4 \xf \x_5    \;,
\ea
where $\sig$ is one of the following combinations
\ba
\sig = &&(-+++),(+-++),(++-+),(+++-), \nn \\
       &&(+---),(-+--),(--+-),(---+)      \;.
	\label{eq-sig1}
\ea
Let us consider $\sig=(-+++)$ as an example.
In this case the solutions are
\ba
	(\gs,\gsp) &=& (\gmu,\gmu\g_5),(\gmu\g_5,\gmu)  \\
	(\xf,\xfp) &=& (\id,\x_5),(\tau_{\mu\nu},
		       \widetilde{\tau}_{\mu\nu}),
		       (\x_5,\id) \;.
\ea
The other solutions can be obtained by applying the
following transformations
\ba
	\gs \longra  \xf^*  &,& \gsp \longra  \xfp^*   \\
	\gs \longra -\gsp &,& \xf  \longra -\xfp   \\
	\gs \longra -\gsp &,& \xf  \longra  \xfp   \;.
\ea
Especially, we are interested in the following solutions
\ba
\label{eq-1tr-dist1}
\o{\gs }{\xf }{} &=& \o{\gmu}{I}{}
\mbox{ and }
\o{\gsp}{\xfp}{} = \o{\gmu\g_5}{\x_5}{}\ ,
\\
\o{\gs }{\xf }{} &=& \o{\gmu}{\x_5}{}
\mbox{ and }
\o{\gsp}{\xfp}{} = \o{\gmu\g_5}{I}{} \ .
\ea
such that the vector and axial channels are either
distance-one or distance-three four-fermion operators.
More generally, as in the two spin trace case,
all solutions are of odd distance (i.e. either distance one or
three). Notice that in the one spin trace formalism
the correct chiral limit is only
obeyed by the sum of both channels including their
chiral partner operators,
but not by each operator independently.
\section{One-Loop Radiative Corrections}
One-loop radiative corrections for the bilinear operators
are calculated in an organized way
in several papers \cite{sheard0,sheard1,sharpe3,Japan1,sharpe4}.
Here we present the important technical steps
in detail to obtain the results for
the radiative corrections for Landau gauge operators.
Most of the diagrams for the radiative corrections for
the four-fermion operators are in common with those
for the bilinear operators.
We will derive the analytic form of
those diagrams which have nothing to do
with the bilinear operators.
There are eight such diagrams: (g1), (g2),
(g3) and (g4) in both Figure 3 and Figure 4.
Let us explain the technical details of calculating
one of the diagrams (Figure 3 (g1)) in a self-contained
way. Since the calculations of other diagrams are quite
similar from the standpoint of technical and
mathematical difficulties.
{}From the Feynman rules given in section 5.1, we can derive
the naive analytic form
of Figure 4 (g1)
\ba
G^{3}_{g1} & = &
\int \overline{dp}
\int \overline{dq}
\int \overline{dk}
\int \overline{dp'}
\int \overline{dq'}
\nonumber \\
& \cdot &
\frac{1}{N_f^2}
\sum_{A,B}\  \exp[i \frac{\pi_{C}}{a} \cdot A a]
\ (\overline{\gamma_S \otimes \xi_F})_{AB}
\  \exp[-i p' \cdot B a]
\nonumber \\
& \cdot &
ia \frac{ \sum_{\nu}
\ \overline{\delta}(p'-p+\pi_{\eta_{\nu}})
\ \sin(p'_{\nu}a) }
{\sum_{\beta} \sin^2(p'_{\beta}a)}
\nonumber \\
& \cdot &
( - i g\  T^I_{\mbox{\small ab}} ) \
\cos( [-\frac{k_{\mu}}{2}- (\frac{\pi_{D}}{a})_{\mu}]a)
\ \overline{\delta}
(p - \frac{\pi_{D}}{a} -k + \frac{\pi_{\eta_{\mu}}}{a})
\nonumber \\
& \cdot &
(-ig\  T^I_{\mbox{\small a}'\mbox{\small b}'} ) \
\cos[(\frac{k_{\xi}}{2} + (\frac{\pi_{C'}}{a})_{\xi})a]
\ \overline{\delta}
(\frac{\pi_{C'}}{a} -q +k + \frac{\pi_{\eta_{\xi}}}{a})
\nonumber \\
& \cdot &
ia \frac
{\sum_{\rho}\  \overline{\delta}(q-q'+\pi_{\eta_{\rho}})
\ \sin(q'_{\rho}a) }
{\sum_{\beta} \sin^2(q_{\beta}a)}
\nonumber \\
& \cdot &
\frac{1}{N^2_f} \sum_{A',B'}
\ \exp[iq'\cdot A'a] \
( \overline{\gamma_{S'} \otimes \xi_{F'}} )_{A'B'}
\ \exp[-i\frac{\pi_{D'}}{a} \cdot B'a]
\nonumber \\
\label{6-g1}
& \cdot &
\left(
\frac{\delta_{\mu\xi}}{ \sum_{\lambda} \frac{4}{a^2}
\sin^2(\frac{k_{\lambda}a}{2})}
-(1-\alpha)
\frac{ \frac{4}{a^2} \sin(\frac{k_{\mu}a}{2})
\sin(\frac{k_{\xi}a}{2})}
{[\sum_{\lambda} \frac{4}{a^2} \sin^2(\frac{k_{\lambda}a}{2})]^2}
\right)
\ea
where
\ba
\int \overline{dk} \equiv
\int_{ -\frac{\pi}{a} }^{ \frac{\pi}{a} } \frac{dk_1}{2 \pi}
\int_{ -\frac{\pi}{a} }^{ \frac{\pi}{a} } \frac{dk_2}{2 \pi}
\int_{ -\frac{\pi}{a} }^{ \frac{\pi}{a} } \frac{dk_3}{2 \pi}
\int_{ -\frac{\pi}{a} }^{ \frac{\pi}{a} } \frac{dk_4}{2 \pi}
\ .
\ea
The following technical relationship is useful to
simplify Eq. (\ref{6-g1}).
\ba
& &
\label{tech-1}
\overline{\delta}(p'-k +
(\frac{\pi_{D}}{a}+ \frac{\pi_{\eta_{\mu}}}{a}
+ \frac{\pi_{\eta_{\nu}}}{a}))
=
\overline{\delta}(p'-k + \frac{\pi_{\tilde{D}''}}{a})
( \hat{\pi}_{\eta_{\nu}} )_{\tilde{D}''\tilde{D}'}
( \hat{\pi}_{\eta_{\mu}} )_{\tilde{D}' D}
\\
& & \mbox{with  } \
\label{tech-2}
(\hat{\pi}_{\eta_{\mu}} )_{\tilde{D}' D} (-1)^{D_{\mu}}
=  (-1)^{\tilde{D}_{\mu}} (\hat{\pi}_{\eta_{\mu}}
)_{\tilde{D}' D}
= (\overline{ \overline{ \gamma_{\mu} \otimes I }}
)_{\tilde{D}' D}
\ea
This relationship is explained in detail in Ref.
\cite{smit1,sheard0,kieu1,sharpe3}.
Using Eq. (\ref{tech-1}, \ref{tech-2}),
we can simplify
Eq. (\ref{6-g1}) as follows
\ba
G^{3}_{g1} & = &
g^2 \sum_{I} T^I_{\mbox{\small ab}}
T^I_{\mbox{\small a}' \mbox{\small b}'}
a^4 \int \overline{dk}
\sum_{\mu}\sum_{\nu}\sum_{\rho}\sum_{\xi}
\frac{1}{N^2_f} \sum_{AB}
\frac{1}{N^2_f} \sum_{A'B'}
\nonumber \\
& \cdot &
\cos(\frac{k_{\mu}a}{2})
\cos(\frac{k_{\xi}a}{2})
\sin( k_{\nu}a) \sin(k_{\rho}a)
F^2(ka)
\nonumber \\
& \cdot &
(\delta_{\mu\xi}B(ka)
-(1-\alpha) 4 \sin(\frac{k_{\mu}a}{2})
\sin(\frac{k_{\xi}a}{2}) B^2(ka) )
\nonumber \\
& \cdot &
(-1)^{C \cdot A}
(\overline{\gamma_S \otimes \xi_F})_{AB}
\exp(-ik \cdot Ba) (-1)^{\tilde{D}'' \cdot B}
(\overline{\overline{\gamma_{\nu\mu}
\otimes I}})_{\tilde{D}'' D}
\nonumber \\
\label{6-g1-1}
& \cdot &
(\overline{\overline{\gamma_{\xi\rho}
\otimes I}})_{C' \tilde{C}''}
(-1)^{\tilde{C}'' \cdot A'}
\exp(ik \cdot A'a)
(\overline{\gamma_{S'} \otimes \xi_{F'}})_{A'B'}
(-1)^{D' \cdot B'}
\ea
In order to make the integral dimensionless, let us
introduce a new variable $ \phi $ such that
$ \phi_{\mu} = k_{\mu}a $.
Also we need the following identities:
\ba
& &
\label{id-1}
\exp(i \phi \cdot A)
(\overline{\gamma_S \otimes \xi_F})_{AB}
 =  \left[\Pi_{\mu} \exp(i \frac{\phi_{\mu}}{2})
\right]
\sum_{M} E_{M}(\phi)
(\overline{\gamma_{MS} \otimes \xi_{MF}})_{AB}
\\
& &
\label{id-2}
(\overline{\gamma_S \otimes \xi_F})_{AB}
\exp(-i \phi \cdot B)
 = \left[\Pi_{\mu} \exp(-i \frac{\phi_{\mu}}{2})
\right]
\sum_{N} E_{N}(-\phi)
(\overline{\gamma_{SN} \otimes \xi_{FN}})_{AB}
\\
& &
\label{id-3}
E_{M} = \Pi_{\mu} \frac{1}{2}
\left[\exp(-i \frac{\phi_{\mu}}{2}) +
(-1)^{\tilde{M}_{\mu}} \exp(+i \frac{\phi_{\mu}}{2})
\right]
\ea
where
\ba
\tilde{M}_{\mu} = \sum_{\nu \ne \mu}
M_{\nu} \ .
\ea
It is useful to introduce the following notations for
the fermion and boson propagators
\ba
F(\phi, m ) & \equiv &
\frac{1}
{\sum_{\alpha} \sin^2( \phi_{\alpha}) + m^2}
\nonumber \\
B(\phi, m ) & \equiv &
\frac{1}
{\sum_{\alpha} 4 \sin^2( \frac{ \phi_{\alpha} }{2})
+ m^2} \ .
\ea
Using Eqs.
(\ref{id-1}, \ref{id-2}, \ref{id-3}) and Eq. (\ref{k-gs})
we can simplify Eq. (\ref{6-g1-1}) further.
\ba
G^{3}_{(g1)} & = &
g^2 \sum_{I} T^I_{\mbox{\small ab}}
T^I_{\mbox{\small a}' \mbox{\small b}'}
\int \frac{d^4\phi}{(2\pi)^4}
\sum_{\mu} \sum_{\nu} \sum_{\rho} \sum_{\xi}
\sum_{M,N}
\nonumber \\
& \cdot &
E_M( \phi) E_N(-\phi)
\cos(\frac{\phi_{\mu}}{2})
\cos(\frac{\phi_{\xi}}{2})
\sin(\phi_{\nu})
\sin(\phi_{\rho}) F^2(\phi,0)
\nonumber \\
& \cdot &
\left[ \delta_{\mu\xi}B(\phi,0)-(1-\alpha)
4 \sin(\frac{\phi_{\mu}}{2})
\sin(\frac{\phi_{\xi}}{2}) B^2(\phi,0) \right]
\nonumber \\
& \cdot &
\label{6-g1-2}
(\overline{\overline{
\gamma_{SN\nu\mu} \otimes \xi_{FN} }})_{CD}
(\overline{\overline{
\gamma_{\xi\rho MS'} \otimes \xi_{MF'} }})_{C'D'}
\ea
Let us note that this expression is divergent
for $ M = N = 0 $.
In order to regularize the infra-red divergences,
we need to put a vanishing mass ($ \kappa a $)
on the gluon propagator.
We need to separate the divergent part
of Eq. (\ref{6-g1-2})
\ba
G^{3}_{(g1)} & = & G^{3,(g1)}_{Convergent} +
G^{3,(g1)}_{Divergent}
\ea
The convergent part of $ G^{3}_{(g1)} $ is
\ba
G^{3,(g1)}_{Convergent} & = &
g^2 \sum_{I} T^I_{\mbox{\small ab}}
T^I_{\mbox{\small a}' \mbox{\small b}'}
\int \frac{d^4\phi}{(2\pi)^4}
\sum_{\mu} \sum_{\nu} \sum_{\rho}
\sum_{M,N}
\nonumber \\
& \cdot &
\left[ E_M( \phi) E_N(-\phi)
\cos^2(\frac{\phi_{\mu}}{2}) \sin(\phi_{\nu})
\sin(\phi_{\rho}) F^2(\phi,0) B(\phi,\kappa a)
\right.
\nonumber \\
& & \left.
- \frac{\delta_{\nu\rho} \delta_{M,0} \delta_{N,0} }
{4} B(\phi,0) B(\phi,\kappa a)\right]
\nonumber \\
& \cdot &
(\overline{\overline{
\gamma_{SN\nu\mu} \otimes \xi_{FN} }})_{CD}
(\overline{\overline{
\gamma_{\mu\rho MS'} \otimes \xi_{MF'} }})_{C'D'}
\nonumber \\
& - & (1-\alpha)
g^2 \sum_{I} T^I_{\mbox{\small ab}}
T^I_{\mbox{\small a}' \mbox{\small b}'}
\int \frac{d^4\phi}{(2\pi)^4}
\sum_{M}
\nonumber \\
& \cdot &
\left[ E_M( \phi) E_M(-\phi) - \delta_{M0}
\right]
B(\phi,0) B(\phi,\kappa a)
\nonumber \\
& \cdot &
(\overline{\overline{
\gamma_{SM} \otimes \xi_{FM} }})_{CD}
(\overline{\overline{
\gamma_{MS'} \otimes \xi_{MF'} }})_{C'D'}
\nonumber \\
& = &
\frac{g^2}{(4\pi)^2} \sum_{I} T^I_{\mbox{\small ab}}
T^I_{\mbox{\small a}' \mbox{\small b}'}
\sum_{\mu} \sum_{\nu} \sum_{\rho}
\sum_{M,N} X^{\mu,\nu\rho}_{MN}
(\overline{\overline{
\gamma_{SN\nu\mu} \otimes \xi_{FN} }})_{CD}
(\overline{ \overline{
\gamma_{\mu\rho MS'} \otimes \xi_{MF'} }})_{C'D'}
\nonumber \\
& - & (1-\alpha)
\frac{g^2}{(4\pi)^2}
\sum_{I} T^I_{\mbox{\small ab}}
T^I_{\mbox{\small a}' \mbox{\small b}'}
\sum_{M} X_{M}
(\overline{\overline{
\gamma_{SM} \otimes \xi_{FM} }})_{CD}
(\overline{ \overline{
\gamma_{MS'} \otimes \xi_{MF'} }})_{C'D'}\ .
\ea
The divergent part of $ G^{3}_{(g1)} $ is
\ba
G^{3,(g1)}_{Divergent} & = &
g^2 \sum_{I} T^I_{\mbox{\small ab}}
T^I_{\mbox{\small a}' \mbox{\small b}'}
\int \frac{d^4\phi}{(2\pi)^4}
B(\phi, 0)B(\phi,\kappa a)
\nonumber \\
& \cdot &
\sum_{\mu} \sum_{\nu}
\frac{1}{4}
(\overline{\overline{
\gamma_{S\nu\mu} \otimes \xi_{F} }}
)_{CD}
(\overline{\overline{
\gamma_{\mu\nu S'} \otimes \xi_{F'} }}
)_{C'D'}
\nonumber \\
& - & (1- \alpha)
g^2 \sum_{I} T^I_{\mbox{\small ab}}
T^I_{\mbox{\small a}' \mbox{\small b}'}
\int \frac{d^4\phi}{(2\pi)^4}
B(\phi, 0)B(\phi,\kappa a)
(\overline{\overline{
\gamma_{S} \otimes \xi_{F} }}
)_{CD}
(\overline{\overline{
\gamma_{S'} \otimes \xi_{F'} }}
)_{C'D'}
\nonumber \\
& = &
\frac{g^2}{(4\pi)^2}
\sum_{I} T^I_{\mbox{\small ab}}
T^I_{\mbox{\small a}' \mbox{\small b}'}
(-\ln(\kappa a)^2 - \gamma_E + F_{0000} + 1)
\nonumber \\
& \cdot &
\left[ \sum_{\mu} \sum_{\nu}
\frac{1}{4}
(\overline{\overline{
\gamma_{S\nu\mu} \otimes \xi_{F} }}
)_{CD}
(\overline{\overline{
\gamma_{\mu\nu S'} \otimes \xi_{F'} }}
)_{C'D'} \right.
\nonumber \\
& - & (1-\alpha) \left.
(\overline{\overline{
\gamma_{S} \otimes \xi_{F} }}
)_{CD}
(\overline{\overline{
\gamma_{S'} \otimes \xi_{F'} }}
)_{C'D'} \right]
\ea
where $ \gamma_E $ is Euler number
and
\ba
F_{0000} & = & (4\pi)^2 \int_{0}^{1}dx x
\exp(-8x) \left[ I_{0}(2x) \right]^4
\nonumber \\
& &
+ (4\pi)^2 \int_{1}^{\infty}dx x
\left[ \exp(-8x) (I_{0}(2x))^4 -
\frac{1}{(4\pi x)^2} \right]
\nonumber \\
& = & 4.3692\cdots
\ea
In the continuum, we have to use the same infra-red
regulator such that both infra-red divergences
($ \ln (\kappa)^2 $)
cancel when we make a connection
between the continuum and lattice operators.
This is true
as long as the anomalous dimensions at one loop
are identical.
The diagrams (g2), (g3) and (g4) in Figure 3
can be calculated analytically in the same
way as (g1) in Figure 4.
The sum of all the contributions of (g1), (g2), (g3) and (g4)
in Figure 3 is
\ba
& & G^3_{(g1)} + G^3_{(g2)}+G^3_{(g3)}+G^3_{(g4)} =
\nonumber \\
& &
-\frac{g^2}{(4\pi)^2}
\sum_{I} T^I_{\mbox{\small ab}}
T^I_{\mbox{\small a}' \mbox{\small b}'}
(-\ln(\kappa a)^2 - \gamma_E + F_{0000} + 1)
\nonumber \\
& & \cdot
\left[ \sum_{\mu} \sum_{\nu}
\frac{1}{4}
(
\overline{\overline{
\gamma_{\mu\nu S} \otimes \xi_{F} }}
-
\overline{\overline{
\gamma_{S\nu\mu} \otimes \xi_{F} }}
)_{CD}
(
\overline{\overline{
\gamma_{\mu\nu S'} \otimes \xi_{F'} }}
-
\overline{\overline{
\gamma_{S'\nu\mu} \otimes \xi_{F'} }}
)_{C'D'} \right]
\nonumber \\
& &-
\frac{g^2}{(4\pi)^2} \sum_{I} T^I_{\mbox{\small ab}}
T^I_{\mbox{\small a}' \mbox{\small b}'}
\sum_{\mu} \sum_{\nu} \sum_{\rho}
\sum_{M,N} X^{\mu,\nu\rho}_{MN}
\nonumber \\
& & \cdot
(
\overline{\overline{
\gamma_{\mu\nu NS} \otimes \xi_{NF} }}
-
\overline{\overline{
\gamma_{SN\nu\mu} \otimes \xi_{FN} }}
)_{CD}
(
\overline{ \overline{
\gamma_{\mu\rho MS'} \otimes \xi_{MF'} }}
-
\overline{ \overline{
\gamma_{S'M\rho\mu} \otimes \xi_{F'M} }}
)_{C'D'}
\nonumber \\
& & + (1-\alpha)
\frac{g^2}{(4\pi)^2}
\sum_{I} T^I_{\mbox{\small ab}}
T^I_{\mbox{\small a}' \mbox{\small b}'}
\sum_{M} X_{M}
\nonumber \\
& & \cdot
(
\overline{\overline{
\gamma_{MS} \otimes \xi_{MF} }}
-
\overline{\overline{
\gamma_{SM} \otimes \xi_{FM} }}
)_{CD}
(
\overline{ \overline{
\gamma_{MS'} \otimes \xi_{MF'} }}
-
\overline{ \overline{
\gamma_{S'M} \otimes \xi_{F'M} }}
)_{C'D'}\ .
\ea

Let us discuss the color one trace diagrams.
The technical calculation is quite similar
to the above procedure for the color two trace diagrams.
The only thing one needs to be careful about is that
the color structure for one color trace diagrams are
different from that for two color trace diagrams.
Performing the technical calculation as complicated as
the above,
the color one trace diagrams
(g1), (g2), (g3) and (g4) in Figure 4
can be expressed analytically as follows
\ba
& & G^4_{(g1)} + G^4_{(g2)}+G^4_{(g3)}+G^4_{(g4)} =
\nonumber \\
& &
\frac{g^2}{(4\pi)^2} (-\ln(\kappa a)^2-\gamma_E + F_{0000} + 1 )
\sum_{\mu} \sum_{\nu} \frac{1}{4}
\nn \\
& & \cdot
\left[
C_F \delta_{\mbox{\small ab}'}
\delta_{\mbox{\small a}' \mbox{\small b}}
\left\{
(\overline{\overline{
\gamma_{\mu\nu S} \otimes \xi_F
}})_{CD}
(\overline{\overline{
\gamma_{S'\nu\mu} \otimes \xi_{F'}
}})_{C'D'}
+
(\overline{\overline{
\gamma_{S\nu\mu} \otimes \xi_{F}
}})_{CD}
(\overline{\overline{
\gamma_{\mu\nu S'} \otimes \xi_{F'}
}})_{C'D'}
\right\}
\right.
\nn \\
& & - \sum_{I} T^I_{\mbox{\small a} \mbox{\small b}'}
T^I_{\mbox{\small a}' \mbox{\small b}}
\left.
\left\{
(\overline{\overline{
\gamma_{\mu\nu S} \otimes \xi_F
}})_{CD}
(\overline{\overline{
\gamma_{\mu\nu S'} \otimes \xi_{F'}
}})_{C'D'}
+
(\overline{\overline{
\gamma_{S\nu\mu} \otimes \xi_{F}
}})_{CD}
(\overline{\overline{
\gamma_{S'\nu\mu} \otimes \xi_{F'}
}})_{C'D'}
\right\}
\right]
\nn \\
& &
-2(1-\alpha)\frac{g^2}{(4\pi)^2}
(-\ln (\kappa a)^2 -\gamma_E + F_{0000} + 1)
(\overline{\overline{
\gamma_{S} \otimes \xi_{F}
}})_{CD}
(\overline{\overline{
\gamma_{S'} \otimes \xi_{F'}
}})_{C'D'}
\nn \\
& & \cdot
\left[
C_F \delta_{\mbox{\small ab}'}
\delta_{\mbox{\small a}' \mbox{\small b}}
- \sum_{I} T^I_{\mbox{\small a} \mbox{\small b}'}
T^I_{\mbox{\small a}' \mbox{\small b}}
\right]
\nonumber \\
& &
+ \frac{g^2}{(4\pi)^2}
\sum_{\mu}\sum_{\nu}\sum_{\rho}\sum_{M,N}
X^{\mu,\nu\rho}_{MN}
\left[
C_F \delta_{\mbox{\small ab}'}
\delta_{\mbox{\small a}' \mbox{\small b}} \right.
\nn \\
& & \cdot
\left\{
(\overline{\overline{
\gamma_{\mu\nu MS} \otimes \xi_{MF}
}})_{CD}
(\overline{\overline{
\gamma_{S'N\rho\mu} \otimes \xi_{F'N}
}})_{C'D'}
+
(\overline{\overline{
\gamma_{SM\nu\mu} \otimes \xi_{FM}
}})_{CD}
(\overline{\overline{
\gamma_{\mu\rho NS'} \otimes \xi_{NF'}
}})_{C'D'}
\right\}
\nn \\
& &
- \sum_{I} T^I_{\mbox{\small a} \mbox{\small b}'}
T^I_{\mbox{\small a}' \mbox{\small b}}
\nn \\
& & \cdot
\left.
\left\{
(\overline{\overline{
\gamma_{\mu\nu MS} \otimes \xi_{MF}
}})_{CD}
(\overline{\overline{
\gamma_{\mu\rho NS'} \otimes \xi_{NF'}
}})_{C'D'}
+
(\overline{\overline{
\gamma_{SM\nu\mu} \otimes \xi_{FM}
}})_{CD}
(\overline{\overline{
\gamma_{S'N\rho\mu} \otimes \xi_{F'N}
}})_{C'D'}
\right\}
\right]
\nn \\
& &
-(1-\alpha)\frac{g^2}{(4\pi)^2}
\sum_{M} X_M
\nn \\
& &
\left[
C_F \delta_{\mbox{\small ab}'}
\delta_{\mbox{\small a}' \mbox{\small b}}
\left\{
(\overline{\overline{
\gamma_{MS} \otimes \xi_{MF}
}})_{CD}
(\overline{\overline{
\gamma_{S'M} \otimes \xi_{F'M}
}})_{C'D'}
+
(\overline{\overline{
\gamma_{SM} \otimes \xi_{FM}
}})_{CD}
(\overline{\overline{
\gamma_{NS'} \otimes \xi_{NF'}
}})_{C'D'}
\right\}\right.
\nn \\
& &
- \sum_{I} T^I_{\mbox{\small a} \mbox{\small b}'}
T^I_{\mbox{\small a}' \mbox{\small b}}
\nn \\
& & \cdot
\left.
\left\{
(\overline{\overline{
\gamma_{MS} \otimes \xi_{MF}
}})_{CD}
(\overline{\overline{
\gamma_{MS'} \otimes \xi_{MF'}
}})_{C'D'}
+
(\overline{\overline{
\gamma_{SM} \otimes \xi_{FM}
}})_{CD}
(\overline{\overline{
\gamma_{S'M} \otimes \xi_{F'M}
}})_{C'D'}
\right\}
\right]
\ea
where
\ba
C_F = \frac{N_c^2 -1}{2 N_c} = \frac{4}{3}
\ea

Let us consider the diagrams in common with those of
the bilinear operators. The color two trace diagrams
(a$\cdots$), (c$\cdots$) and (d$\cdots$) in Figure 3
are identical to those of the bilinear operators.
The color one trace diagrams
(a$\cdots$), (c$\cdots$) and (d$\cdots$)
in Figure 4 for the Landau gauge operators are
equivalent to those of the bilinear operators
as long as
the color structure is treated carefully.
The details of the one-loop radiative correction for
the bilinear operators are explained in Ref.
\cite{sheard0,sharpe3,Japan1}.
The results are
\ba
G^3_{(a1)} & = &
\frac{g^2}{(4\pi)} C_F
\delta_{\mbox{\small ab}}
\delta_{\mbox{\small a}' \mbox{\small b}'}
\left[(\sigma_S - (1 - \alpha))
(-\ln(\kappa a)^2 -\gamma_E + F_{0000} + 1)
\right.
\nn \\
& & \cdot
(\overline{\overline{
\gamma_{S} \otimes \xi_{F}
}})_{CD}
(\overline{\overline{
\gamma_{S'} \otimes \xi_{F'}
}})_{C'D'}
\nn \\
& &
+\sum_{\mu}\sum_{\nu}\sum_{\rho}\sum_{M,N}
X^{\mu,\nu\rho}_{MN}
(\overline{\overline{
\gamma_{\mu\nu MSN \rho\mu} \otimes \xi_{F}
}})_{CD}
(\overline{\overline{
\gamma_{S'} \otimes \xi_{F'}
}})_{C'D'}
\nn \\
& & \left.
- (1-\alpha) \sum_M X_M
(\overline{\overline{
\gamma_{MSM} \otimes \xi_{MFM}
}})_{CD}
(\overline{\overline{
\gamma_{S'} \otimes \xi_{F'}
}})_{C'D'} \right]
\\
G^4_{(a1)} & = &
\frac{g^2}{(4\pi)}
\sum_{I} T^I_{\mbox{\small a} \mbox{\small b}'}
T^I_{\mbox{\small a}' \mbox{\small b}}
\left[(\sigma_S - (1 - \alpha))
(-\ln(\kappa a)^2 -\gamma_E + F_{0000} + 1)
\right.
\nn \\
& & \cdot
(\overline{\overline{
\gamma_{S} \otimes \xi_{F}
}})_{CD}
(\overline{\overline{
\gamma_{S'} \otimes \xi_{F'}
}})_{C'D'}
\nn \\
& &
+\sum_{\mu}\sum_{\nu}\sum_{\rho}\sum_{M,N}
X^{\mu,\nu\rho}_{MN}
(\overline{\overline{
\gamma_{\mu\nu MSN \rho\mu} \otimes \xi_{F}
}})_{CD}
(\overline{\overline{
\gamma_{S'} \otimes \xi_{F'}
}})_{C'D'}
\nn \\
& & \left.
- (1-\alpha) \sum_M X_M
(\overline{\overline{
\gamma_{MSM} \otimes \xi_{MFM}
}})_{CD}
(\overline{\overline{
\gamma_{S'} \otimes \xi_{F'}
}})_{C'D'} \right]
\\
G^3_{(c1)} & = &
\frac{g^2}{(4\pi)} C_F
\delta_{\mbox{\small ab}}
\delta_{\mbox{\small a}' \mbox{\small b}'}
\left[
-\alpha(-\ln(\kappa a)^2 -\gamma_E + F_{0000} + 1)
-\frac{1}{8} Z_{0000} - R
\right]
\nn \\
& & \cdot
(\overline{\overline{
\gamma_{S} \otimes \xi_{F}
}})_{CD}
(\overline{\overline{
\gamma_{S'} \otimes \xi_{F'}
}})_{C'D'}
\\
G^4_{(c1)} & = &
\frac{g^2}{(4\pi)} C_F
\delta_{\mbox{\small a} \mbox{\small b}'}
\delta_{\mbox{\small a}' \mbox{\small b}}
\left[
-\alpha(-\ln(\kappa a)^2 -\gamma_E + F_{0000} + 1)
-\frac{1}{8} Z_{0000} - R
\right]
\nn \\
& & \cdot
(\overline{\overline{
\gamma_{S} \otimes \xi_{F}
}})_{CD}
(\overline{\overline{
\gamma_{S'} \otimes \xi_{F'}
}})_{C'D'}
\\
G^3_{(d1)} & = &
\frac{g^2}{(4\pi)} C_F
\delta_{\mbox{\small ab}}
\delta_{\mbox{\small a}' \mbox{\small b}'}
\frac{1}{8}(3+\alpha)Z_{0000}
(\overline{\overline{
\gamma_{S} \otimes \xi_{F}
}})_{CD}
(\overline{\overline{
\gamma_{S'} \otimes \xi_{F'}
}})_{C'D'}
\\
G^4_{(d1)} & = &
\frac{g^2}{(4\pi)} C_F
\delta_{\mbox{\small a} \mbox{\small b}'}
\delta_{\mbox{\small a}' \mbox{\small b}}
\frac{1}{8}(3+\alpha)Z_{0000}
(\overline{\overline{
\gamma_{S} \otimes \xi_{F}
}})_{CD}
(\overline{\overline{
\gamma_{S'} \otimes \xi_{F'}
}})_{C'D'}
\ea
where
\ba
Z_{0000} & = & (4\pi)^2\int \frac{d^4\phi}{(2\pi)^4}
B(\phi,0) = (4\pi)^2 \int_{0}^{\infty} dx
\exp(-8x) [I_{0}(2x)]^4 = 24.46604
\\
R & = & (4\pi)^2 \int \frac{d^4\phi}{(2\pi)^4}
\left[ \cos(\phi_1)
\left((2\sin^2(\phi_1)-\frac{1}{F(\phi)}\right)
\right.
\nn \\
& & \cdot
\left.
\left(-2 -2 \sin^2(\frac{\phi}{2}) + \frac{1}{4B}
\right) B(\phi,0) F^2(\phi) -
B^2(\phi,0)\right]
\nn\\
& = & -5.2145\cdots
\ea
and we follow the same notation as in Ref. \cite{Japan1}.

The diagrams (c1) and (d1) in Figures 3 and Figure 4 were first
calculated in Feynman gauge in Ref. \cite{smit1}.
The analytic expression of the diagram (a2) can be
obtained from that of (a1) in Figure 3 and Figure 4
simply by switching the bilinear indices.
The analytic expression of the diagrams (c2), (c3) and
(c4) are identical to that of (c1)
in Figure 3 and Figure 4.
Also the analytic expression of the diagrams (d2), (d3) and
(d4) are identical to that of (d1)
in Figure 3 and Figure 4.

Now we have the analytic expression of all the
diagrams necessary for the Landau gauge operator.
Let us introduce the following definitions
\ba
G^3_{(g)} & \equiv & G^3_{(g1)} + G^3_{(g2)} +
G^3_{(g3)} + G^3_{(g4)}
\\
G^4_{(g)} & \equiv & G^4_{(g1)} + G^4_{(g2)} +
G^4_{(g3)} + G^4_{(g4)}
\\
G^3_{(a,c,d)} & \equiv &
G^3_{(a1)} + G^3_{(a2)} +
2 (G^3_{(c1)}+G^3_{(d1)})
\\
G^4_{(a,c,d)} & \equiv &
G^4_{(a1)} + G^4_{(a2)} +
2 (G^4_{(c1)}+G^4_{(d1)}) \ .
\ea
Note that only half of the self-energy diagrams are taken
into consideration since the other half corresponds to the
the wave function renormalization of the external fields.

Now we need the following useful
SU(3) Fierz transformation identity
\ba
\label{su3-id}
\sum_{I} T^I_{\mbox{\small ab}}
T^I_{\mbox{\small a}' \mbox{\small b}'}
= -\frac{1}{2N_c}
\delta_{\mbox{\small ab}}
\delta_{\mbox{\small a}' \mbox{\small b}'}
+\delta_{\mbox{\small a} \mbox{\small b}'}
\delta_{\mbox{\small a}' \mbox{\small b}}\ .
\ea
Using Eq. (\ref{su3-id}),
for $ (\overline{\overline{
\gamma_S \otimes \xi_F
}}) = (\overline{\overline{
\gamma_{S'} \otimes \xi_{F'}
}}) = (\overline{\overline{
\gamma_{\mu} \otimes \xi_5 }}) $
we find
\ba
G^3_{(g)} & = & - \frac{g^2}{(4\pi)^2}
(-\ln(\kappa a)^2 -\gamma_E + F_{0000} + 1)
\left[ -(A \times P)_{ab;a'b'} + 3 (A \times P)_{ab';a'b}
\right]
\nn \\
& &
- 4 \frac{g^2}{(4\pi)^2} X^{1,22}_{(0000),(0000)}
\left[
-(A \times P)_{ab;a'b'} + 3 (A \times P)_{ab';a'b}
\right]
\nn \\
& &
- 4 \frac{g^2}{(4\pi)^2}
(4 X^{1,11}_{(1111),(1111)} + 6 X^{1,22}_{(1111),(1111)} )
\left[ -\frac{1}{6}(A \times S)_{ab;a'b'}
+ \frac{1}{2} (A \times S)_{ab';a'b} \right]
\nn \\
& &
+ 4 (1-\alpha)\frac{g^2}{(4\pi)^2} X_{(1111)}
\left[ -\frac{1}{6}(A \times S)_{ab;a'b'}
+ \frac{1}{2} (A \times S)_{ab';a'b} \right]
+ \cdots
\\
G^4_{(g)} & = &
\frac{g^2}{(4\pi)^2} \left[
(3+2\alpha) (-\ln(\kappa a)^2 -\gamma_E + F_{0000} + 1)
+8 X^{1,11}_{(0000),(0000)} + 12 X^{1,22}_{(0000),(0000)}
\right.
\nn \\
& & \left.
-2(1-\alpha)X_{(0000)} \right]
\cdot
\left[ \frac{3}{2}(V \times P)_{ab';a'b}
-\frac{1}{2} (V \times P)_{ab;a'b'} \right]
\nn \\
& &
-12 \frac{g^2}{(4\pi)^2}
\left[\frac{1}{4}(-\ln(\kappa a)^2 -\gamma_E + F_{0000} + 1)
+ X^{1,22}_{(0000),(0000)} \right]
\nn \\
& & \cdot
[ \frac{7}{6} (A \times P)_{ab';a'b}
+ \frac{1}{2} (A \times P)_{ab;a'b'}]
\nn \\
& &
- \frac{g^2}{(4\pi)^2}
\left[4 X^{1,11}_{(1111),(1111)}
+ 6 X^{1,22}_{(1111),(1111)}
-(1-\alpha) X_{(1111)} \right]
\nn \\
& & \cdot
\left[ -\frac{7}{3}(A \times S)_{ab';a'b}
- (A \times S)_{ab;a'b'} \right]
\nn \\
& &
-\frac{g^2}{(4\pi)^2}
6 X^{1,22}_{(1111),(1111)}
\left[3 (V \times S)_{ab';a'b}
- (V \times S)_{ab;a'b'} \right]
+ \cdots
\\
G^3_{(a,c,d)} & = & \frac{g^2}{(4\pi)^2}
(28.706\cdots) (V \times P)_{ab;a'b'}
\\
G^4_{(a,c,d)} & = & \frac{g^2}{(4\pi)^2}
\left[(30.40479\cdots)(V \times P)_{ab';a'b}
-(0.56626\cdots)(V \times P)_{ab;a'b'}) \right]
\ea
where
\ba
& & X^{1,22}_{(0000),(0000)} = -0.822800\cdots
\\
& & X^{1,11}_{(1111),(1111)} = 0.041042\cdots
\\
& & X^{1,22}_{(1111),(1111)} = 0.023793\cdots
\\
& & X_{(1111)} = 0.076312\cdots
\ea
{}From the above equations, we finally obtain
\ba
(V \times P)^{(1)}_{ab;ba} & = &
37.446 (V \times P)^{(0)}_{ab;ba}
- 2.9136 (V \times P)^{(0)}_{aa;bb}
- 5.25285 (A \times P )^{(0)}_{ab;ba}
\nonumber
\\
& &
- 2.2512 (A \times P)^{(0)}_{aa;bb}
-0.5381 ( A \times S)^{(0)}_{ab;ba}
-0.2306 ( A \times S)^{(0)}_{aa;bb}
\nonumber
\\
& &
+0.4283 ( V \times S)^{(0)}_{ab;ba}
-0.1428 ( V \times S)^{(0)}_{aa;bb}
+ \cdots
\\
(V \times P)^{(1)}_{aa;bb} & = &
+ 28.706  (V \times P)^{(0)}_{aa;bb}
- 4.5024 (A \times P)^{(0)}_{ab;ba}
+ 1.5008 (A \times P)^{(0)}_{aa;bb}
\nonumber
\\
& &
-0.4612 (A \times S)^{(0)}_{ab;ba}
+0.1537 (A \times S)^{(0)}_{aa;bb}
+ \cdots \ ,
\ea
where the anomalous dimension terms and
momentum-dependent terms are neglected,
and
the superscript represents the number of loops.
The anomalous dimension matrix can be obtained
simply by reading off the coefficients of the terms
proportional to $ \ln (\kappa a) $.

The results for the $ (A \times P) $ channel
can be obtained through a procedure similar to
the above for the $ (V \times P) $ channel:
\ba
(A \times P)^{(1)}_{ab;ba} & = &
-5.25285 (V \times P)^{(0)}_{ab;ba}
-2.2512  (V \times P)^{(0)}_{aa;bb}
+ 37.976 (A \times P)^{(0)}_{ab;ba}
\nonumber
\\
& &
- 4.5043 (A \times P)^{(0)}_{aa;bb}
+ 0.4283 (A \times S)^{(0)}_{ab;ba}
- 0.1428 (A \times S)^{(0)}_{aa;bb}
\nonumber
\\
& &
- 0.5381 (V \times S)^{(0)}_{ab;ba}
- 0.2306 (V \times S)^{(0)}_{aa;bb}
+\cdots
\\
(A \times P)^{(1)}_{aa;bb} & = &
-4.5024 (V \times P)^{(0)}_{ab;ba}
+1.5008 (V \times P)^{(0)}_{aa;bb}
+ 24.464 (A \times P)^{(0)}_{aa;bb}
\nonumber
\\
& &
- 0.4612 (V \times S)^{(0)}_{ab;ba}
+ 0.1537 (V \times S)^{(0)}_{aa;bb}
+\cdots  \ ,
\ea
where the anomalous dimension terms and
momentum-dependent terms are neglected,
and
the superscript represents the number of loops.

The scheme-dependent finite mixing
of the chiral partner operators (
$ (A \times S)_{ab;ba} $, $ (A \times S)_{aa;bb} $,
$ (V \times S)_{ab;ba} $ and $ (V \times S)_{aa;bb} $)
in Eqs. (\ref{chi-1})--(\ref{chi-4})
can be obtained by multiplying both bilinears in
the above equations
by $ ( \overline{\gamma_5 \otimes \xi_5}) $.
\newpage
\section{Figure Caption}
\begin{figure}[h]
\caption[1]{Feynman Rules: (a) gluon propagator,
(b) fermion propagator, (c) one gluon vertex,
(d) two gluon vertex, (e) operator with no gluon emitted,
(f) operator with one gluon emitted,
(g) operator with two gluon emitted.
The dashed lines in (e), (f) and (g)
represents gauge link between the quark and anti-quark fields
as well as flow of color indices.}
\caption[2]{Feynman Diagrams of one loop correction
for the bilinear operators}
\caption[3]{Feynman Diagrams of one loop correction
for four-fermion operators in color two trace form}
\caption[4]{Feynman Diagrams of one loop correction
for four-fermion operators in color one trace form}
\end{figure}
\ed
\newpage
\begin{thebibliography}{99}
%
\bibitem{b-lee} M. K. Gaillard and B. Lee,
Phys. Rev. D 10, (1974) 897.
%
\bibitem{wise1} M. Wise, CALT-68-1518,
Lectures delivered at Banff Summer Instute on Particle
and Fields, Banff, Canada, Aug. 14-27 1988, p. 124.
%
\bibitem{wise2} F. J. Gilman and M. B. Wise, Phys. Rev. D27
(1983) 1128.
%
\bibitem{wise3} F. J. Gilman and M. B. Wise, Phys. Rev. D20
(1979) 2392.
%
\bibitem{paschos} E. A. Paschos and U. T\"{u}rke,
Phys. Rep. 178, (1989) 145.
%
%
\bibitem{sharpe0} S. R. Sharpe, A. Patel, R. Gupta,
G. Guralnik and G. W. Kilcup, Nucl. Phys. B286 (1987)
253.
%
\bibitem{sharpe1}
Stephen R. Sharpe, {\em Staggered Fermions on
the Lattice and $ \cdots $} appears in
{\bf Standard Model, Hadron
Phenomenology and Weak Decays on the Lattice}, DOE/ER/40614-5.
%
\bibitem{fuku}
N. Ishizuka, et al., Phys. Rev. Lett. Vol. 71,
(1993) 24.
%
\bibitem{kilcup0}
G. Kilcup, S. Sharpe, R. Gupta and A. Patel,
Phys. Rev. Lett. Vol. 64, (1990) 25.
%
\bibitem{smit1} M. F. L. Golterman nad J. Smit,
Nucl. Phys. B245 (1984) 61.
%
\bibitem{klubergstern1} H. Kluberg-stern, A Morel,
O. Napoly, Nucl. Phys. B220 (1983) 447.
%
\bibitem{kieu1} D.Daniel and T. D. Kieu, Phys. Lett. B175
(1986) 73.
%
\bibitem{sheard0} D. Daniel and S. N. Sheard, Nucl. Phys.
B302 (1988) 471.
%
\bibitem{sheard1} S. N. Sheard, Nucl. Phys.
B314 (1989) 238.
%
\bibitem{daniel1} D. Daniel, S. Hands, T. D. Kieu and
S. N. Sheard, Phys. Lett. B193 (1987) 85.
%
\bibitem{tool-kit}
Kilcup, G. and Sharpe, S., Nucl. Phys. B283,
(1987) 493.
%
\bibitem{buras} A. Buras and P. Weisz, Nucl. Phys. B333
(1990) 69.
%
\bibitem{alti} G. Altarelli, G. Curci, G. Martinelli,
S. Petrarca, Nucl. Phys. B187 (1981) 461.
%
\bibitem{collins} J. Collins, {\em Renormalization}
(Cambridge University Press, Cambridge, England, 1984).
%
\bibitem{sharpe3} A. Patel and S. R. Sharpe, Nucl. Phys.
B395 (1993) 701.
%
\bibitem{Japan1} N. Ishizuka and Y. Shizawa, UTHEP-261.
Phys. Rev. D (to be published).
%
\bibitem{sharpe4} S. R. Sharpe and A. Patel,
Nucl. Phys. B417 (1994) 307.
%
\bibitem{verstegen} D. Verstegen, Nucl. Phys. B249 (1985)
685.
%
\bibitem{mandula} J. Mandula, Nucl. Phys. B228 (1983)
91.
%
\bibitem{los} The authors thank Dr. S. Sharpe for
pointing this difficulty and suggesting
this resolution. The authors thank Dr. Rajan Gupta for his
hospitality at the Santa Fe meeting where
we benefited from conversations with Dr. S. Sharpe.
Los Alamos National Laboratory held
Santa Fe Workshop organized by Dr. Rajan Gupta between
July 23, 1994 and August 13, 1994.
%
\bibitem{parisi} G. Parisi, in
{\it High Energy Physics--1980
(XX International Conference, Madison, Wisconsin)},
Proceedings of the XX International Conference
on High Energy Physics, edited by L. Durand and
L.G. Pondrom, AIP Conf. Proc. No. 68
(AIP, New York, 1981) 1531.
%
\bibitem{lepage} P. Lepage and P. Mackenzie,
Phys. Rev. D48 (1993) 2250.
%
\bibitem{karsch} U. Heller and F. Karsch, Nucl. Phys.
B251 (1985) 254.
%
\bibitem{wlee} W. Lee, Phys. Rev. D49 (1994) 3563.
%
%
\end{thebibliography}
